\documentclass{article}
\pdfoutput=1
\usepackage[affil-it]{authblk}
\usepackage{graphicx}
\usepackage[space]{grffile}
\usepackage{latexsym}
\usepackage{amsfonts,amsmath,amssymb}
\usepackage{url}
\usepackage[utf8]{inputenc}
\usepackage{textcomp}
\usepackage{longtable}
\usepackage{jheppub}

\usepackage{tikz}
\usetikzlibrary{%
    decorations.pathreplacing,%
    decorations.pathmorphing%
}
\usepackage{mathtools}
\usetikzlibrary{calc}

\begin{document}

\title{Prompt neutrinos and intrinsic charm at SHiP}

\author{Weidong Bai and Mary Hall Reno}
\affiliation{Department of Physics and Astronomy, University of Iowa, Iowa City, IA 52242, USA}

\emailAdd{weidong-bai@uiowa.edu}
\emailAdd{mary-hall-reno@uiowa.edu}
\date{\today}

\abstract{{We present a new evaluation of the far-forward neutrino plus antineutrino flux and number of events
from charm hadron decays in a 400 GeV proton beam dump experiment like the Search for Hidden Particles (SHiP).
Using next-to-leading order perturbative QCD and a model for intrinsic charm, we include intrinsic transverse
momentum effects and other kinematic angular corrections. We compare this flux to a far-forward flux evaluated with
next-to-leading order perturbative QCD, without intrinsic transverse momentum, that used the angular distribution of charm quarks rather than the
neutrinos from their decays.
The tau neutrino plus antineutrino number of events in the perturbative QCD evaluation 
is reduced by a factor of about three when intrinsic transverse momentum and the full decay 
kinematics are included. 
We show that intrinsic charm contributions can significantly enhance the number of events from neutrinos from charm hadron decays. Measurements of the  number of events from tau neutrino plus antineutrino interactions and of the muon charge asymmetry as a function of energy can be used to constrain intrinsic charm models.}
}

\maketitle
\flushbottom

\section{Introduction}

The potential for the discovery of dark matter and hidden particles in theories beyond the standard model are the motivation for new beam dump experiments under review or design. The SHiP (Search for Hidden Particles) experiment is
one such project, now in the comprehensive design phase \cite{Anelli:2015pba,Alekhin:2015byh}. The experiment plans to run with $2\times 10 ^{20}$ protons on target over a 5-year run using the SPS 400 GeV proton beam at CERN. While a major goal is to detect or constrain physics beyond the standard model, the protons incident on the beam dump target will produce a significant number of neutrinos. Designed to minimize the production of pions and kaons to optimize the beam dump as a source of hidden particles, guaranteed outputs are neutrinos from charm {hadrons} that decay promptly, as well as some neutrinos from pion and kaon decays. The
short lifetimes of charm {hadrons} mean that they decay before losing energy by hadronic scattering in the target. 

Of particular interest are tau neutrinos, as the direct detection of tau neutrinos is thus far limited to 9 $\nu_\tau$ events at DONUT \cite{Kodama:2000mp,Kodama:2007aa} and 10 tau neutrino candidate events at Opera \cite{Pupilli:2016zrb,Agafonova:2018auq}. In ref. \cite{Alekhin:2015byh}, an evaluation of the tau neutrino flux with some simplifying approximations was made,  predicting on the order of 1,000 tau neutrino and antineutrino interactions in the SHiP detector. Muon neutrinos and antineutrinos are also produced from charm {hadron} decays.
{The charged and neutral $D$ mesons decay semileptonically to muon neutrinos and electron neutrinos.}
 Cabibbo and kinematic suppression factors mean that the $D^\pm$ mesons decay to $\nu_\tau$ in two-body decays {with a branching fraction $<1.2\times 10^{-3}$ \cite{Patrignani:2016xqp}}, and the $D^0,\bar{D}^0$, not at all. The $D_s^\pm$ is heavy enough, and not suppressed by the Cabibbo angle in its decay, so that
$D_s^-\to \tau \bar{\nu}_\tau$ and its charge conjugate have a branching fraction of $(5.48\pm 0.23)\%$ \cite{Patrignani:2016xqp}.
 Subsequent tau decays, also prompt, produce a second tau neutrino for each tau antineutrino that comes directly from the $D_s^-$.
{Overall, perturbative production of charm results in a flux of muon neutrinos and antineutrinos at a level that is a factor of 
$\sim 10$ larger than the flux of tau neutrinos and antineutrinos.}

The SHiP experimental configuration, with a {molybdenum target} beam dump and $51.5$m distance between the front of the target and the detector, collects particles in the very forward direction, for a narrow range of angles $\theta$ relative to the beam direction. For the nominal SHiP detector, angles $\theta < 7.3\times 10^{-3}-1.9\times 10^{-2}$ rad reflect the width and height of the detector $2{\rm m}\times 
0.75{\rm m}$, respectively. In terms of pseudorapidity, this means $\eta>
4.6-5.6$. 
In ref. \cite{Alekhin:2015byh}, the neutrino flux was evaluated in a collinear approximation using
next-to-leading order (NLO) {perturbative} QCD for the $pA\to c\bar{c} X$ cross section. In the collinear approximation used there, the charm quark momentum direction was used to determine $\eta_c$, which was used to approximate whether or not the neutrinos and antineutrinos from charm decays were directed into the detector.
After applying a charm momentum angular (or pseudorapidity) cut, the energies of the mesons, taus and eventually neutrinos were traced using analytic energy distributions where relative angles are already integrated.

{In this paper, our primary focus is a refined evaluation of the  flux of
neutrinos and antineutrinos at SHiP that includes the angular dependence of charm hadron decays to neutrinos and intrinsic transverse momentum effects in charm hadron production. We evaluate the forward flux of tau neutrinos
plus antineutrinos for SHiP using NLO perturbative QCD production of charm to compare to the evaluation in ref. \cite{Alekhin:2015byh}. 
In this paper, we also evaluate the flux of neutrinos using a model of intrinsic charm.
Models of enhanced charm production (see, e.g., refs. \cite{Duraes:1995hy,Hobbs:2013bia,Maciula:2017wov}) including intrinsic charm \cite{Brodsky:1980pb,Brodsky:1981se,Vogt:1994zf,Gutierrez:1998bc,Navarra:1995rq,Dulat:2013hea,Jimenez-Delgado:2014zga,Hou:2017khm,Carvalho:2017zge}
have been explored in the context of the high energy atmospheric neutrino flux
\cite{Halzen:2016pwl,Halzen:2016thi,Laha:2016dri,Giannini:2018utr,Bhattacharya:2018tbc}. 
Using the model of intrinsic charm described in refs. \cite{Brodsky:1980pb,Brodsky:1981se,Vogt:1994zf,Gutierrez:1998bc} and normalized according 
to ref. \cite{Laha:2016dri} for their evaluation of the atmospheric neutrino flux, we find a significant increase in the number of events from tau neutrinos and antineutrinos at SHiP.}

{Intrinsic charm introduces particle-antiparticle asymmetries
at SHiP that we evaluate here for muon neutrino and antineutrino events.
Gluon fusion,
$g g\to c\bar{c}X$, is the dominant parton model process in perturbative production of charm \cite{Nason:1987xz,Nason:1989zy,Mangano:1991jk,Cacciari:1998it,Cacciari:2001td}, so particle-antiparticle asymmetries are not significant. On the other hand, for five  or seven particle states (e.g., $|uudc\bar{c}\rangle$,
$|uudc\bar{c}u\bar{u}\rangle$) where
valence quark momenta are brought to, for example, 
$\overline{D}^0=u\bar{c}$, but not a $D^0 = \bar{u}c$, particle-antiparticle asymmetries can be large. They are significant for $D^+/D^-$ and $D^0/\bar{D}^0$, but considerably less significant for $D_s^+/D_s^-$.  Thus, asymmetries in the muon neutrino and antineutrino fluxes, but not the tau neutrino and antineutrino fluxes, are predicted with intrinsic charm.
We make
an approximate evaluation of the muon neutrino-antineutrino asymmetries relying on our perturbative charm evaluation, intrinsic charm and an approximate flux of muon neutrinos and antineutrinos at SHiP from pion and kaon decays
based on fluxes evaluated in ref. \cite{Anelli:2015pba}.
 }

{We begin in sec. 2 with the production of charm quarks and hadrons in the parton model and with a model for intrinsic charm based on refs. \cite{Brodsky:1980pb,Brodsky:1981se,Vogt:1994zf} discussed in detail in ref. \cite{Gutierrez:1998bc}. 
Our assumptions about hadronization and the incorporation of intrinsic transverse momentum are included in sec. 2, where we compare our NLO
perturbative QCD transverse momentum distribution of $D^++D^-+D^0+\bar{D}^0$  in $pp$ collisions with $E_p=400$ GeV 
to LEBC-EHS measurements \cite{AguilarBenitez:1988sb}.
We show energy and rapidity
distributions of charm hadrons to illustrate some of the features of charm production that translate through charm hadron decays to the neutrino energy and 
pseudorapidity distributions, at least approximately. Efforts to measure production of charm hadrons
with the 400 GeV beam at CERN, e.g., the DsTau project \cite{Aoki:2017spj}, would help refine the perturbative calculation of the neutrino flux and 
constrain forward production of charm hadrons {from
intrinsic charm} directly.

In sec. 3, we show our results for neutrino energy distributions from prompt decays of charm hadrons produced by a beam of 400 GeV protons directed to the SHiP detector. 
Given $N_p=2\times 10^{20}$ protons and a SHiP detector represented by a column depth of 66.7 g/cm$^2$ of lead, we evaluate the number of events.
As a proxy for detector angular cuts, we restrict pseudorapidity values  to $\eta>5.3$. When the cut is applied to the charm pseudorapidity ($\eta_c>5.3$), with the same parameter inputs as in ref. \cite{Alekhin:2015byh}, we find the number of 
tau neutrino plus antineutrino events is 10\% lower than the number of events determined using the more detailed detector geometry of ref. \cite{Alekhin:2015byh}.
Given this good agreement, all of the neutrino and
antineutrino results presented here approximate the detector geometry by requiring $\eta_\nu>5.3$.
The primary impact of the calculational improvements discussed here is to reduce the number of neutrinos incident on  the SHiP detector. We discuss the
impact of each correction {and their associated uncertainties} in sec. 3.

We also discuss how an enhanced number of tau neutrino and antineutrino events and their energy dependence 
could show evidence of intrinsic charm. As noted above, intrinsic
charm models predict particle-antiparticle asymmetries that translate to different muon neutrino and antineutrino fluxes.
Sec. 3 also includes an evaluation of the asymmetry in the number of 
muon neutrino and antineutrino events that shows how intrinsic charm changes the perturbative prediction for charm. With an approximate
number of muon neutrinos and antineutrinos from pion and kaon decays, we show the muon charge asymmetry at SHiP as a function of energy.
The discussion in sec. 4 summarizes our results for the number of events for tau neutrinos and antineutrinos, and the muon charge asymmetry,
at SHiP. We point to uncertainties in our evaluations and to where
refinements to the calculation can be made.

Relevant formulas for intrinsic charm are collected in appendix A.
Appendix B gives some details of the tau neutrino energy distributions that come from the decays of the $D_s$ directly to $\nu_\tau$ and the
chain decay $D_s\to\tau\to \nu_\tau$.
Our
treatment of the 
muon neutrino and antineutrino fluxes from pion and kaon decays needed for the muon charge asymmetry, based on the results of ref. 
\cite{Anelli:2015pba}, appears in appendix C.

}

\section{Production of charm quarks and hadrons}

Charm quark production in perturbation theory is evaluated {using the collinear parton model}.
The factorization scale $\mu_F$ characterizes this collinear approximation, with the interpretation that, roughly, partons with transverse momenta below $\mu_F$ are approximately collinear.
At SHiP, because of the small angular size of the detector downstream of the target, even small transverse momentum effects are potentially important. We discuss below first, the standard next-to-leading order evaluation of perturbative charm production.
Then, we show our implementation of intrinsic transverse momentum to model the low
transverse momentum of the partons and parameters of quark fragmentation to hadrons. Finally, we show how intrinsic charm could contribute to charm hadron production
 at SHiP.
 
 \subsection{Perturbative charm}
 
The parton model evaluation of charm production involves the production of a charm quark pair, followed by the hadronization of the quark  and antiquark 
into a charm meson or baryon. Charm production can be written in terms of the parton level hard
scattering cross section $\hat{\sigma}_{ij}$ that depends on the strong coupling constant 
$\alpha_s(Q^2)$
evaluated at a renormalization scale $Q^2=\mu_R^2$. The differential cross section at the parton level is convoluted 
with the parton distributions for partons $i$ and $j$ via the parton distribution functions $f_i^{H_1}(x_1,\mu_F^2)$ and
$f_j^{H_2}(x_2,\mu_F^2)$, with $x_{1,2}$ equal to the momentum fractions of the participating partons for hadronic collisions 
$H_1+H_2\to c\bar{c} X$. At leading order in QCD, 
gluon fusion dominates the production of $c\bar{c}$ pairs, as it does at next-to-leading order (NLO)
 \cite{Nason:1987xz,Nason:1989zy,Mangano:1991jk,Cacciari:1998it,Cacciari:2001td}.
  
The differential cross section for the inclusive $c$ cross section for charm momentum $p$ is
\begin{eqnarray}
\nonumber
E\frac{d^{3}\sigma}{d\, p^{3}} & =&\sum_{i,j=q,\overline{q},g}\int dx_{1}dx_{2}f_{i}^{H_1}(x_{1},\mu_{F}^{2})f_{j}^{H_2}(x_{2},\mu_{F}^{2})\nonumber \\
 & \times& \left[E\frac{d^{3}\widehat{\sigma}_{ij}(x_{1}P_{H_1},x_{2}P_{H_2},p,m^{2},\mu_{F}^{2},\mu_{R}^{2})}{d\, p^{3}}\right]\ ,
\label{eq:crossection}
\end{eqnarray}
for parton cross section $\hat{\sigma}$. {Implicit in eq. (\ref{eq:crossection}) are integrals over the $\bar{c}$ momentum for the leading-order
and NLO $2\to2$ processes, and over the $\bar{c}$ and quark or gluon momentum in the $2\to 3$ processes. Explicit expressions for the one-particle inclusive heavy quark differential cross section, after these integrals are performed, appear in ref. \cite{Nason:1989zy}.}

At SHiP, a proton beam, $H_1=p$, is incident on a molybdenum target, so $H_2=$Mo. The molybdenum atomic
mass is $A=96$ {and charge $Z=42$. The cross section for charm production is dominated by $gg$ interactions which are independent of the number of neutrons and protons. There are potential valence quark effects in the forward production of charm, however.} We approximate the parton distribution function (PDF) for partons in the {nucleus} by
\begin{equation}
\label{eq:fMo}
f_j^{\rm Mo}(x,\mu_F^2)=Z f_j^{p/A}(x,\mu_F^2) + (A-Z) f_j^{n/A}(x,\mu_F^2)\ 
\end{equation}
for charm production, {\rm where we assume isospin symmetry to relate the up and down quark distributions in the neutron to those in the proton.}

Theoretical evaluations of perturbative production of charm have a long history, with evaluations through NLO in QCD in refs. \cite{Nason:1987xz,Nason:1989zy,Mangano:1991jk} that have been implemented in the HVQ computer program
{with the HVQMNRPHO main program and supporting fortran subroutines}. 
Our perturbative evaluation uses the HVQ computer program as a basis for the NLO evaluation. {We use the single inclusive charm quark evaluation provided there, with the addition of intrinsic transverse momentum and an implementation of  the $D_s$ decay $D_s\to \tau \nu_\tau$ and 
subsequent tau decays, as described below.}
For the main results presented here, we take the charm quark mass to be $m_c=1.27$ GeV, and the central values of the factorization
scale $\mu_F=2.1 m_c$ and renormalization scale $\mu_R=1.6 m_c$, following ref. \cite{nelson2013narrowing} where
$m_c=1.27\pm 0.09$ GeV is used. The scale choices {with $m_c=1.27$ GeV permit us to make a direct comparison with the results in ref. 
 \cite{Alekhin:2015byh}. With $m_c=1.27$ GeV, we can also take advantage of the detailed data-constrained analysis of charm production over a wide energy range performed in ref. \cite{nelson2013narrowing}
 to set the range of scales to be $\mu_F=1.25-4.65\, m_c$ and $\mu_R=1.5-1.7\, m_c$.}

 We use the nCTEQ15 PDFs
\cite{Kovarik:2015cma}, {available for both free and bound nucleons.
The nCTEQ15 distribution of grids does not include Mo PDFs, but the PDFs for Kr ($A=84$) and Ag ($A=108$) give essentially the same $c\bar{c}$ cross section for $E_p=400$ GeV, where we use eq. (\ref{eq:fMo}) for Mo. 

Before we discuss the nuclear PDF corrections, we note that the charm pair production cross section  per nucleon at next to leading order with 
$\mu_F=2.1\, m_c$, $\mu_R=1.6\, m_c$ and $m_c=1.27$ GeV
 for a proton
beam energy of $E_p=400$ GeV incident on a Mo target is 16.3 $\mu$b using free nucleon nCTEQ15 PDFs, whether one includes the
specific $Z$ and $A$ dependence for Mo or uses an isoscalar nucleon target. 
The nCTEQ15 free nucleon PDFs give a charm pair production cross section that is 14\% lower than the cross section using the
 CT14 PDFs \cite{Dulat:2015mca} for the same charm quark mass and scale choices.
 The measured cross section for $pp$ scattering with $E_p=400$ GeV 
is $18.1\pm 1.7$ $\mu$b as determined  by Lourenco and Wohri in ref. \cite{Lourenco:2006vw}
from data reported in ref. \cite{AguilarBenitez:1988sb}. For our range of renormalization and factorization
scales, the $pp$ cross section with a 400 GeV beam energy is $\sigma_{c\bar{c}}=13.3-19.1$ $\mu$b, with the lower cross section
within $3\sigma$ and a central value of 16.2 $\mu$b that is about 12\% lower than the central value of the measured $pp$ charm pair cross section at this energy.

In general, it is not the case that the upper and lower theoretical bounds on the cross section are obtained
by taking {both the lower renormalization and factorization scales, or both  the upper scale choices}.  However, for $E_p=400$ GeV, it is the case. At this energy, the lower renormalization scale gives a larger cross section because of a larger $\alpha_s$. 
Larger factorization scales result in  PDF evolution that pushes the PDFs to smaller $x$ values. For $E_p=400$ GeV, the larger factorization scale reduces the PDFs in the $x$ region of interest. At higher incident beam energies, the evolved enhanced small $x$ PDF distributions are more important than they are for $E_p=400$ GeV.
 
 Distributions are often evaluated using the transverse mass, $m_T=\sqrt{p_T^2
 + m_c^2}$. In our single inclusive distributions, we can evaluate the cross section with this scale choice. With $m_c=1.27$ GeV, using $\mu_F=2.1 m_T$
 and $\mu_R= 1.6 m_T$, the $pp$ cross section  is $\sigma_{c\bar{c}}=13.3\ \mu$b, on the lower side of the cross section range.
In the results presented here, we use a default scale dependence on the charm quark mass rather than the transverse mass. Below, we use data from ref. \cite{AguilarBenitez:1988sb} to set the parameter in a Gaussian intrinsic transverse momentum distribution. A different parameter for intrinsic transverse momentum would be obtained if the transverse mass scale is used instead. 

At high energies where very small $x$ PDFs are important, nuclear shadowing can have a large impact on forward production, even for
fairly low $A$ targets like nitrogen, as shown in the context of the atmospheric neutrino flux from charm \cite{Bhattacharya:2016jce}. At the
energy considered here, there are two largely compensating effects: the shadowing at smaller $x$ values
and the EMC effect at large $x$ are offset by anti-shadowing for  $x$ in the range of $\sim 0.01-0.3$
(see, e.g., refs. \cite{Dulat:2015mca,Eskola:2016oht,Eskola:2009uj} and references therein). 
The  net effect is that charm pair cross section is enhanced with nuclear corrections. The cross section is 18.5 $\mu$b for bound nucleons in Mo using the central PDF of nCTEQ15 for Ag, so an increase of about 13\% due to nuclear corrections. By varying the nuclear PDFs over the 33 sets of nCTEQ15, the cross section on Mo ranges from 17.8 $\mu$b--19.4 $\mu$b, about $\pm 5\%$.
For reference, the $gg$ initial state accounts for 94\% of the cross section on bound nucleons for the central PDF choice in molybdenum for $E_p=400$ GeV.
 }

The charm hadron and anti-hadron production rates are {essentially equal for perturbative production of charm}. This will distinguish
perturbative charm production from intrinsic charm production, discussed below.

\subsection{Intrinsic transverse momentum and fragmentation}

{For the large rapidities required by SHiP, intrinsic transverse momentum can deflect charm quarks from the detector direction, so it should be included. 
The fragmentation of charm quarks to charm hadrons also influences the transverse momentum distributions, as was noted, for example, already in ref. \cite{Frixione:1994nb} and more recently for LHC energies in ref. \cite{Vogt:2018oje}.   
The intrinsic transverse momentum depends on $\sqrt{s}$ \cite{nelson2013narrowing}. The LEBC-EHS experiment \cite{AguilarBenitez:1988sb} with 400 GeV protons incident on a hydrogen
target is our benchmark for setting the intrinsic transverse momentum distribution value given the fragmentation functions used here.
Our default values for fragmentation parameters,
the same as in ref. \cite{Alekhin:2015byh},  allow us to compare the intrinsic transverse momentum improved evaluation with the collinear 
approximated evaluation \cite{Alekhin:2015byh}.}

{
For hadrons $H$ equal to charged and neutral $D$'s and for $\Lambda_c$, we can write
\begin{equation}
D_{c\to H}(z,\mu_0^2)=\frac{N_H\, z(1-z)^2}{[(1-z)^2+\epsilon_H z]^2}\ ,
\end{equation}
according to the Peterson fragmentation form \cite{Peterson:1982ak}, where we take $z$ as the ratio of the momentum of the meson 
to the momentum of the charm quark, $\vec{p}_H=z \vec{p}_c$, in the lab frame. We use $\epsilon_H$ from 
Kniehl and Kramer \cite{Kniehl:2006mw}, with $\mu_0=1.5$ GeV, and we rescale $N_H$ to normalize the fragmentation functions
to the fragmentation fractions of 
ref. \cite{Lisovyi:2016qjn}. The parameters $\epsilon_H$ and the fragmentation fractions shown in table \ref{table:fragmentation} are used as default values for the results shown here. We do not evolve
the fragmentation functions. The Peterson fragmentation functions are used  for all of the distributions
of  charm hadrons and neutrinos evaluated using NLO perturbative QCD presented here, unless specifically noted.  }

Schweitzer, Teckentrup and Metz \cite{Schweitzer:2010tt} showed that a Gaussian model of intrinsic transverse momentum is compatible with semi-inclusive deep-inelastic scattering and with Drell-Yan production. 
In the Gaussian approach, azimuthal symmetry around the beam axis permits us to write
\begin{equation}
f_{k_T}(\vec{k}_T) = \frac{1}{\pi\langle k_T^2\rangle}\exp \Biggl(-\frac{k_T^2}{\langle k_T^2\rangle}\Biggr)\ ,
\end{equation}
normalized to unity with the integral over $d^2k_T$.  For the Gaussian distribution,
\begin{equation}
\langle k_T^2\rangle = \frac{4}{\pi}\langle k_T\rangle ^2\ .
\end{equation}
The differential distribution for charm
quark production becomes
\begin{equation}
E \frac{d\sigma}{dp_z\, d^2 p_T}= \int d^2 k_T \int d^2 p_T '\, f_{k_T}(\vec{k}_T) E \frac{d\sigma }{dp_z\, d^2 p_T'}
\delta^2 (\vec{p}_T- \vec{p}_T \,'-\vec{k}_T)\ .
\end{equation}
In ref. \cite{Schweitzer:2010tt}, it was shown that for leading order Drell-Yan production,
the Gaussian approximation 
provides a good model if
\begin{equation}
\langle k_{T,DY}^2\rangle  = 1.4\ {\rm GeV}^2
\end{equation}
for a 300 GeV proton beam scattering on platinum (FNAL-288) \cite{Ito:1980ev}. For pion scattering from tungsten (FNAL E615), the value is a little larger (1.7 GeV$^2$) \cite{Palestini:1985zc}.

\begin{figure}
\begin{center}
\hspace{-0.85cm}
\includegraphics[width=0.75\textwidth]{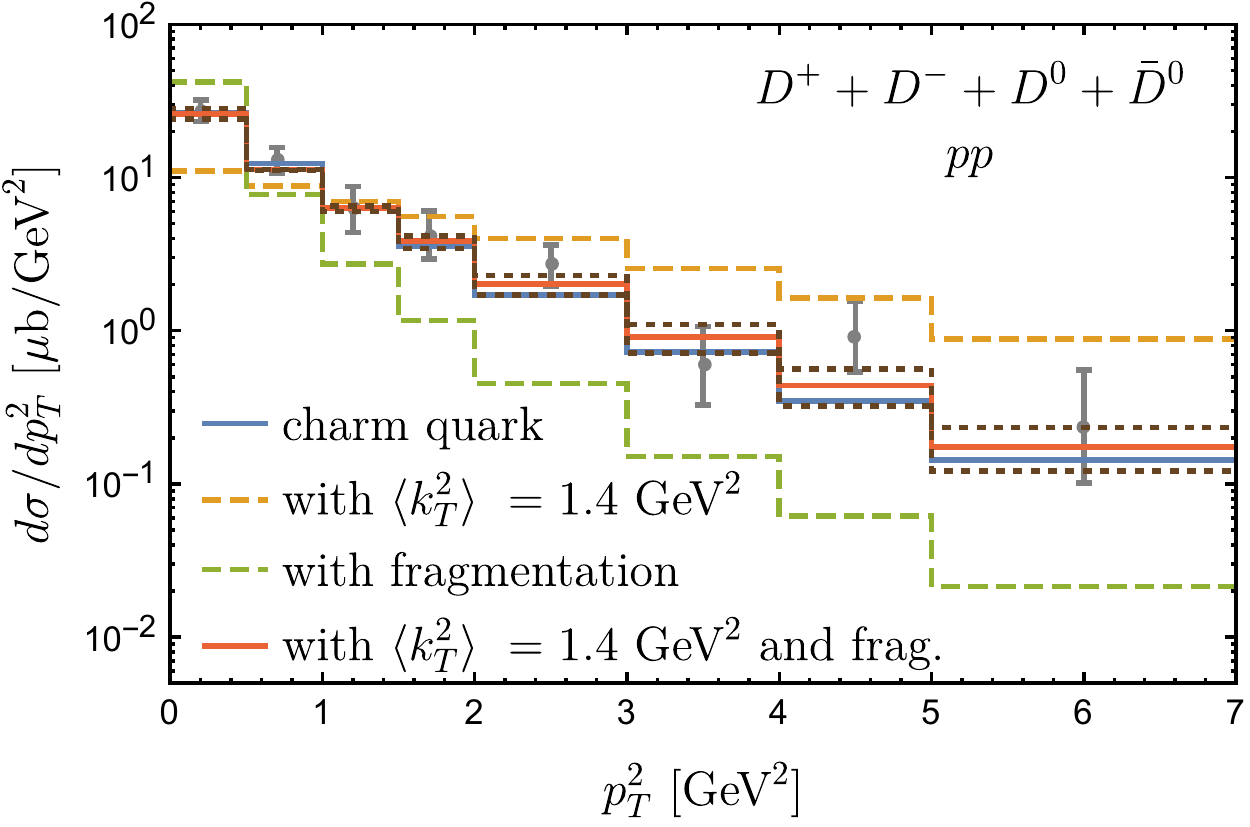}
\caption{The NLO perturbative  differential cross section for $pp$ interactions for $E_p=400$ GeV on a hydrogen target ($\sqrt{s}=27.4$ GeV) to produce
$D^++D^-+D^0+\bar{D}^0$,
as a function of the charm hadron $p_T^2$
with intrinsic transverse momentum $\langle k_T^2\rangle=1.4$ GeV$^2$ (solid red histogram) 
and $\epsilon_D=0.10$ (see text). The dashed black histograms
show the range of $\langle k_T^2\rangle =1.1-1.7$ GeV$^2$, including fragmentation. The orange dashed histogram shows the $p_T^2$ distribution of the charm quark with $\langle k_T^2\rangle = 1.4$ GeV$^2$ (no fragmentation), and the green dashed histogram shows the $D$ meson
$p_T^2$ distribution calculated with $\langle k_T^2\rangle =0$. The solid blue histogram shows the result without fragmentation and without intrinsic transverse momentum.
The data are from the LEBC-EHS experiment \cite{AguilarBenitez:1988sb}.}
\label{fig:lebc}
\end{center}
\end{figure}

{Using the  Peterson fragmentation function for $c\to D^\pm,\ D^0 $ and $\bar{D}^0$ with $\epsilon_D=0.10$, our NLO evaluations with $\langle k_T^2\rangle=1.1-1.7$ GeV$^2$ provide good fits with the $D$ meson $p_T^2$ distribution measured by the LEBC-EHS group \cite{AguilarBenitez:1988sb}. The best fit with the data has
$\langle k_T^2\rangle =1.4$ GeV$^2$. 
The best fit  distribution (solid red histogram) and the range of predictions for $\langle k_T^2\rangle =1.1-1.7$ GeV$^2$ (dashed black histograms)
with the LEBC-EHS data are shown in fig. \ref{fig:lebc}. } 
We find a similar range of $\langle k_T^2\rangle$ values when we compare our predictions with the charm meson data from a 250 GeV proton beam incident on
copper \cite{Alves:1996qz} when we make a 10\% correction to the overall normalization of the $p_T^2$ distribution. 
{To illustrate the separate fragmentation and intrinsic transverse momentum effects, we also show in fig. \ref{fig:lebc} the
charm distribution without  fragmentation or intrinsic transverse momentum (solid blue histogram), and with intrinsic transverse momentum but without fragmentation (dashed orange histogram). The green dashed histogram shows the result of including fragmentation, but not intrinsic transverse momentum.
}

{In the evaluations of charm hadron production and the resulting neutrino energy and rapidity distributions, 
we use} $\langle k_T^2\rangle = 1.4$ GeV$^2$ for the average
intrinsic transverse momentum squared. This is equivalent to $\langle k_T\rangle\simeq 1.05$ GeV. 
We discuss the impact of the range of acceptable 
values of $\langle k_T^2\rangle$ on our predicted number of tau neutrino plus antineutrino events at SHiP in sec.  4.

{The choice of $\epsilon _D=0.10$ for $D^0$ and $D^+$ production is larger than  a more typical choice of $\epsilon_D=0.06$. With our default value, we can make direct comparisons with the muon neutrino results evaluated in the collinear approximation in ref. \cite{Alekhin:2015byh}. As noted above, the choice of  $\langle k_T^2\rangle$
depends on the fragmentation function. With $\epsilon_D=0.06$, $\langle k_T^2\rangle = 0.9-1.4$ GeV$^2$ gives a similar band of predictions to the choice of  $\epsilon_D=0.10$, $\langle k_T^2\rangle = 1.1-1.7$ GeV$^2$. Intrinsic transverse momentum choices outside the respective ranges of values are problematic for the larger values of $p_T^2$ shown in fig. \ref{fig:lebc}, despite the large error bars that span a factor of $\sim 2$ in the highest $p_T^2$ bin. The large errors in the highest $p_T^2$ bins are not reflected proportionately in the error estimates for the neutrino events because the detector angular size is small.
At low $p_T^2$, the experimental error bars are of order $\pm 15\%$, while the spread in predictions with this fragmentation, for the range of $\langle k_T^2\rangle$, is less than $\pm 10\%$. }

\begin{table}[tb]
\begin{center}
\begin{tabular}{|l|c|c|c|}
\hline
Particle &   $\epsilon_H$ & $B_{c\to H}$ \\
\hline
$D^0$  & 0.101 & 0.606\\
\hline
$D^+$   & 0.104 & 0.245\\
\hline
$D_s^+$  & 0.0322  & 0.079\\
\hline
$\Lambda_c^+ $  & 0.00418 & 0.062\\
\hline
\end{tabular}
\caption{\label{table:fragmentation} 
Parameters for the charm quark fragmentation to $H=D$ mesons and the $H=\Lambda_c$, from ref. \cite{Kniehl:2006mw} table I, and the fragmentation fractions of ref. \cite{Lisovyi:2016qjn}}
\end{center}
\vspace{-0.6cm}
\end{table}

\subsection{Intrinsic charm}

Beam dump experiments have the potential to constrain additional contributions to charm hadron production, beyond perturbation theory, through neutrino measurements.  The description of high momentum charm hadrons have long been a topic of interest,  e.g.,  as discussed in refs. \cite{Duraes:1995hy,Hobbs:2013bia,Maciula:2017wov,Brodsky:1980pb,Brodsky:1981se,Vogt:1994zf,Gutierrez:1998bc,Navarra:1995rq,Dulat:2013hea,Jimenez-Delgado:2014zga,Hou:2017khm,Carvalho:2017zge}.
Often cited are the data from the LEBC-MPS collaboration from 800 GeV protons incident on a bubble chamber \cite{Ammar:1988ta}, and the SELEX (E781) collaboration's measurements of high momentum $\Lambda_c$'s in $\pi^-$, $\Sigma^-$ and $p$ beams incident on copper and carbon targets \cite{Garcia:2001xj}. These data show an excess of events for high momentum fractions, beyond the perturbative prediction in this energy regime, however, there are large error bars. They also found asymmetries in the production of hadrons with components that could come from beam valence quarks compared with their antiparticles. 

For definiteness to find charm contributions to neutrino fluxes at SHiP, we use an intrinsic charm model of Brodsky, Hoyer, Peterson and Sakai (BHPS) \cite{Brodsky:1980pb,Brodsky:1981se}, described in
detail in refs. \cite{Vogt:1994zf,Vogt:1995fsa,Gutierrez:1998bc}.
There,  intrinsic charm (IC) is modeled by considering  protons with 5-quark and 7-quark Fock states in which two of the quarks are a $c\bar{c}$ pair.
Heavy and light quarks can 
coalesce {or heavy quarks can independently fragment} into high momentum charm hadrons. While this is not
the dominant charm hadron production mechanism, it {may} dominate the large Feynman $x_F$ distribution. {Our goal here is not
to provide a definitive model of IC. We use this model to illustrate the potential for SHiP to explore the 
contributions of IC, without introducing specific fragmentation functions, binding energy and mass effects beyond what is outlined below.}

Based on frame-independent probability distributions for intrinsic charm $P_{ic}$, two types of contributions
to the production of charm hadrons are included: independent (uncorrelated) fragmentation (F), and
coalescence distributions (C) specific to individual charm hadrons. 
They depend on the $n$-particle probability functions  \cite{Gutierrez:1998bc},
\begin{equation}
\frac{dP_{ic}^n}{dx_1... dx_n} = N_n\alpha_s(M_{c\bar{c}})\frac{\delta (1-\sum _{i=1}^n x_i)}{(m_p^2-\sum_{i=1}^n(\hat{m}_i/x_i))^2}
\end{equation}
for $n=5$ and $n=7$. The uncorrelated fragmentation is assumed to have a delta function fragmentation function for $c\to H$, so as in ref. \cite{Gutierrez:1998bc},
\begin{equation}
\label{eq:frag}
\frac{dP_{ic}^{nF}}{dx_H} = \int dx_1\ .\ .\ .\  d x_n\frac{dP_{ic}^n}{dx_1\  . \ .\ .\  dx_n}\delta (x_H-x_c)\ .
\end{equation}
For coalescence, the probability function includes a delta function that combines the momentum fractions of the
quarks  that 
make up the final state hadron, 
\begin{equation}
\label{eq:coal}
\frac{dP_{ic}^{nC}}{dx_H} = \int dx_1\ .\ .\ .\  d x_n\frac{dP_{ic}^n}{dx_1\  . \ .\ .\  dx_n}\delta (x_H-x_{H_{1V}}- \ .\ . \ .\ - x_{H_{nV}})\ .
\end{equation}
In eqs. (\ref{eq:frag}) and (\ref{eq:coal}), $x_H$ is the Feynman $x_F$ variable for the given hadron $H$. In the large $x_F$ limit, $x_F\simeq x_E$ where,
for SHiP, $x_E= E_H^{\rm lab}/400$ GeV. The Feynman $x_F$ distribution
from intrinsic charm becomes
\begin{equation}
\frac{d\sigma_{ic}(pN)}{dx_F}= \sigma_{pN}^{\rm in}\frac{\mu^2}{4\hat{m}_c^2}\sum_{n=5,7}
\Biggl(\frac{dP_{ic}^{nF}}{dx_F} + \frac{dP_{ic}^{nC}}{dx_F}\Biggr)\ ,
\end{equation}
where $ \sigma_{pN}^{\rm in}$ is the inelastic $pN$ cross section and $\mu^2/4 \hat{m}_c^2$ sets the scale of the intrinsic charm cross section.
In what follows, we approximate $x_F\simeq x_E$ in all of our evaluations of the neutrinos from intrinsic charm.

{Alternatives to the delta functions in eqs. (\ref{eq:frag}) and (\ref{eq:coal}) can be used to relate 
the quark momentum to the hadron momentum, for example, the Peterson fragmentation function
can be used for the uncorrelated fragmentation. We use the delta functions 
for intrinsic charm to be consistent in how we include the uncorrelated and
coalescence contributions. There are many more uncertainties in the theoretical treatment of intrinsic charm than for perturbative production of
charm. A more detailed treatment of fragmentation, binding energy, mass effects and other approximations in this
BHPS model of intrinsic charm would be required should evidence of intrinsic charm
be observed experimentally.}

The relative contributions of the five quark $(uudc\bar{c})$ and seven quark $(uudc\bar{c}q\bar{q})$
terms for $q\bar{q}=u\bar{u},\ d\bar{d},\ s\bar{s}$ are set according to ref. \cite{Gutierrez:1998bc}. The normalizations are
$P_{ic}^5= 3.1\times 10^{-3}$, and $P_{icu}^7=P_{icd}^7=(\hat{m}_c/\hat{m}_u)^2\, P_{ic}^7$, $P_{ics}^7=(\hat{m}_c/\hat{m}_s)^2 \, P_{ic}^7$  where 
$P_{ic}^7=4.4\times 10^{-2}\, P_{ic}^{5}$ and $\hat{m}_u=\hat{m}_d=0.45$ GeV, $\hat{m}_s=0.71$ GeV and $\hat{m}_c=1.8$ GeV. For completeness, we write the differential distributions for the probabilities for each of the charm hadrons as a function of 
$x_F$ in the appendix.

The overall normalization of intrinsic charm remains to be fixed, the value of which is of great interest 
in the context of the prompt atmospheric flux \cite{Laha:2016dri,Halzen:2016thi,Halzen:2016pwl}. Charm production in the atmosphere, followed by its prompt
decay into leptonic or semileptonic decay modes, is the dominant contribution to the atmospheric 
neutrino flux at sufficiently high energies \cite{Enberg:2008te,Bhattacharya:2015jpa,Gauld:2015kvh,Bhattacharya:2016jce,Garzelli:2016xmx,Benzke:2017yjn,Goncalves:2018zzf,Halzen:2016pwl,Halzen:2016thi,Laha:2016dri,Giannini:2018utr,Bhattacharya:2018tbc}. 
The IceCube measurements of the diffuse astrophysical flux have ruled out an
extremely large intrinsic charm component, as discussed in, e.g., ref. \cite{Halzen:2016thi}. Nevertheless, intrinsic charm may play an important role in the high energy atmospheric neutrino flux because the incident cosmic ray hadrons have a very steep energy dependence. Laha and Brodsky in ref. \cite{Laha:2016dri} argue that IceCube will be able to constrain intrinsic charm in the future via their neutrino measurements in the 100 TeV to PeV energy range. 
Laha and Brodsky's normalization of the intrinsic charm cross section 
is based on the highest $x_F$ data in $pp$ collisions for $E_b=800$ GeV ($\sqrt{s}=39$ GeV) measured by the LEBC-MPS collaboration \cite{Ammar:1988ta}.
Using  this highest bin at $x_F=0.32$, Laha and Brodsky choose \cite{Laha:2016dri}:
\begin{eqnarray}
\label{eq:ic25}
\frac{d\sigma_{ic}(pN \to \sum D^\pm,D^0,\bar{D}^0)}{dx_F}\Biggl|_{x_F=0.32} &=&  25\ \mu{\rm b}\quad\quad {\rm (IC25)}\\ 
\frac{d\sigma_{ic}(pA)}{dx_F} = A^{\alpha} \frac{d\sigma_{ic}(pN)}{dx_F}\ , & &
\label{eq:icnorm}
\end{eqnarray}
{a value above the central value in the bin, but within the error bar \cite{Ammar:1988ta}.}

We vary the overall IC normalization in our discussion of the potential signals
of  intrinsic charm
{in the neutrino signal, for example, considering a normalization of the $pN$ differential cross section at $x_F=0.32$,}
\begin{equation}
\label{eq:ic7}
\frac{d\sigma_{ic}(pN \to \sum D^\pm,D^0,\bar{D}^0)}{dx_F}\Biggl|_{x_F=0.32} =  7\ \mu{\rm b}\quad\quad {\rm (IC7)}\ ,
\end{equation} 
also considered
in ref. \cite{Laha:2016dri}. {The IC7 normalization is such that the differential cross section is at the lower end of the LEBC-MPS error bar in the $x_F=0.32$ bin for $E_b=800$ GeV \cite{Ammar:1988ta}. }

The intrinsic charm cross section in this approach is expected to scale with energy as the $pp$ total cross section does, so {it depends only weakly on} energy. We use
these same normalizations (IC25 and IC7) for our evaluation of the intrinsic charm contribution to the neutrino flux
at SHiP {where $E_p=400$ GeV. 
Intrinsic charm added with the IC25 normalization is still within the error bars for the lowest and highest $p_T^2$ shown in
fig. \ref{fig:lebc}, but the fit to the data with $\langle k_T^2\rangle=1.4$ GeV$^2$ is not as good as with the perturbative contribution alone. A larger
intrinsic charm normalization would over-produce $D$ mesons in the lowest $p_T^2$ bin. The sum of the IC7 plus perturbative production of $D$ mesons as a function of $p_T^2$ has a similar fit to
the data shown in fig. \ref{fig:lebc}.}

The weak energy dependence of intrinsic charm means that the relative contributions of intrinsic to perturbative charm production is larger at beam energies of 400 GeV than at the LHC {or for the prompt atmospheric flux at IceCube. The lower beam energy at SHiP measurements is an advantage for constraining the intrinsic charm component.}

To account for nuclear effects, we use the appropriate scaling of $A^\alpha$ where $\alpha=0.71$ \cite{Badier:1983dg}. 
For molybdenum, $A^{0.71}/A=0.27$. We note that a change in the value of $\alpha$ can be absorbed in a change to the overall normalization of the intrinsic charm contribution. For example, $\alpha=0.67$ gives an IC contribution that is $\sim 80\%$ of the IC contribution with $\alpha=0.71$. Experimental evidence of intrinsic charm would require a more detailed study of its $A$ dependence.

Asymmetries
between $D^-$ and $D^+$, and between $D^0$ and $\bar{D}^0$, for the $x_F\simeq x_E$ distributions are predicted in intrinsic
charm models.  Asymmetries are smaller for $D_s^\pm$. The detailed probability distributions listed in the appendix quantify the differences in particle and antiparticle
production as a function of $x_F$. 

In this paper, we will also explore the implications for asymmetries in the high energy
$\nu_\mu$ and $\bar{\nu}_\mu$ energy distributions. Neutrino and antineutrino production from pion and kaon decays are also asymmetric, but the beam dump will suppress pion and kaon production of muon neutrinos and antineutrinos at high energies. We make an approximate evaluation of the muon neutrino and antineutrino flux from pion and kaon decay for the SHiP target configuration
based on fluxes presented in ref.  \cite{Anelli:2015pba}, with normalizations outlined in appendix C. We compare them with contributions from intrinsic and perturbative charm.

\subsection{Charm hadron production}

In this section, we begin with our results for charm hadron production via the perturbative approach for $E_p=400$ GeV. 
Since we are interested in the neutrino and antineutrino energy distributions, most of the figures in this section display the charm hadron
energy distributions to show the impact of elements of the calculation that will eventually translate to the lepton energy distributions.
{They also serve as a reference for measurements of charm hadron production, e.g., the DsTau project \cite{Aoki:2017spj}.}

\begin{figure}
\begin{center}
\hspace{-0.85cm}
\includegraphics[width=0.75\textwidth]{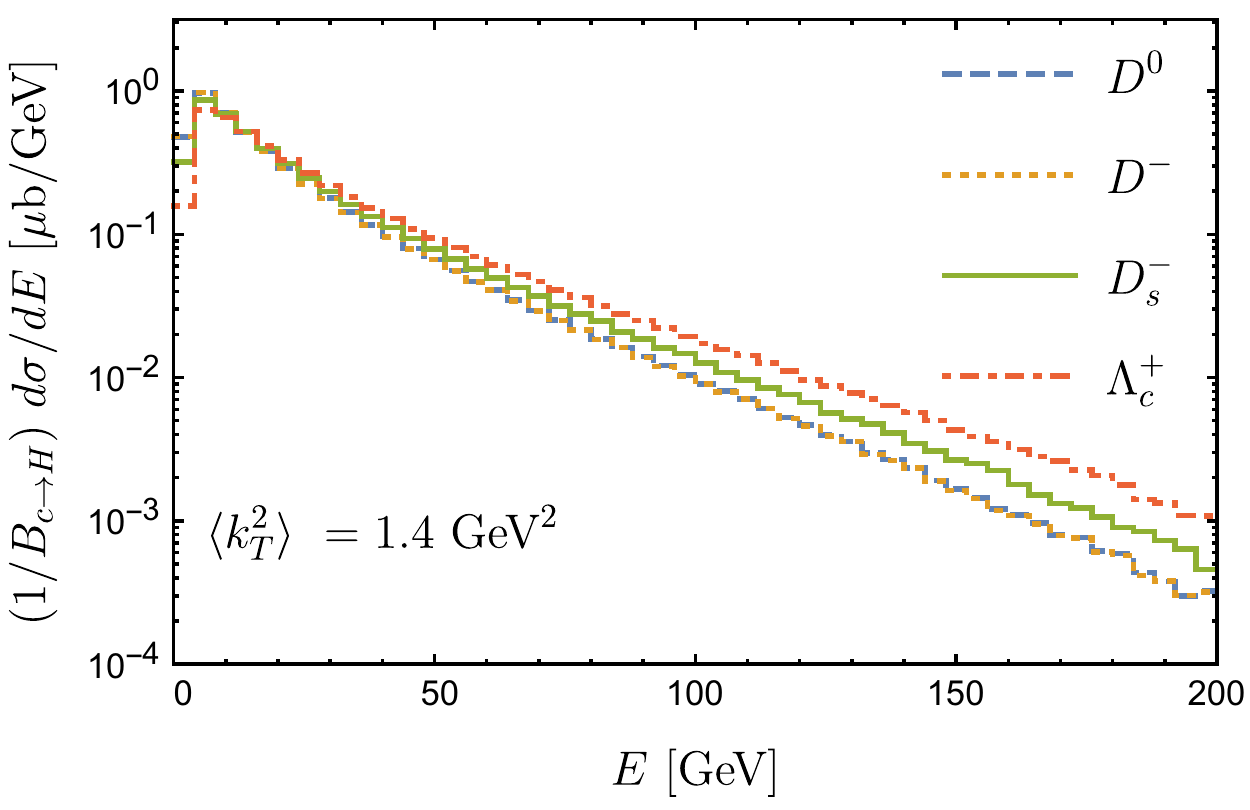}
\includegraphics[width=0.75\textwidth]{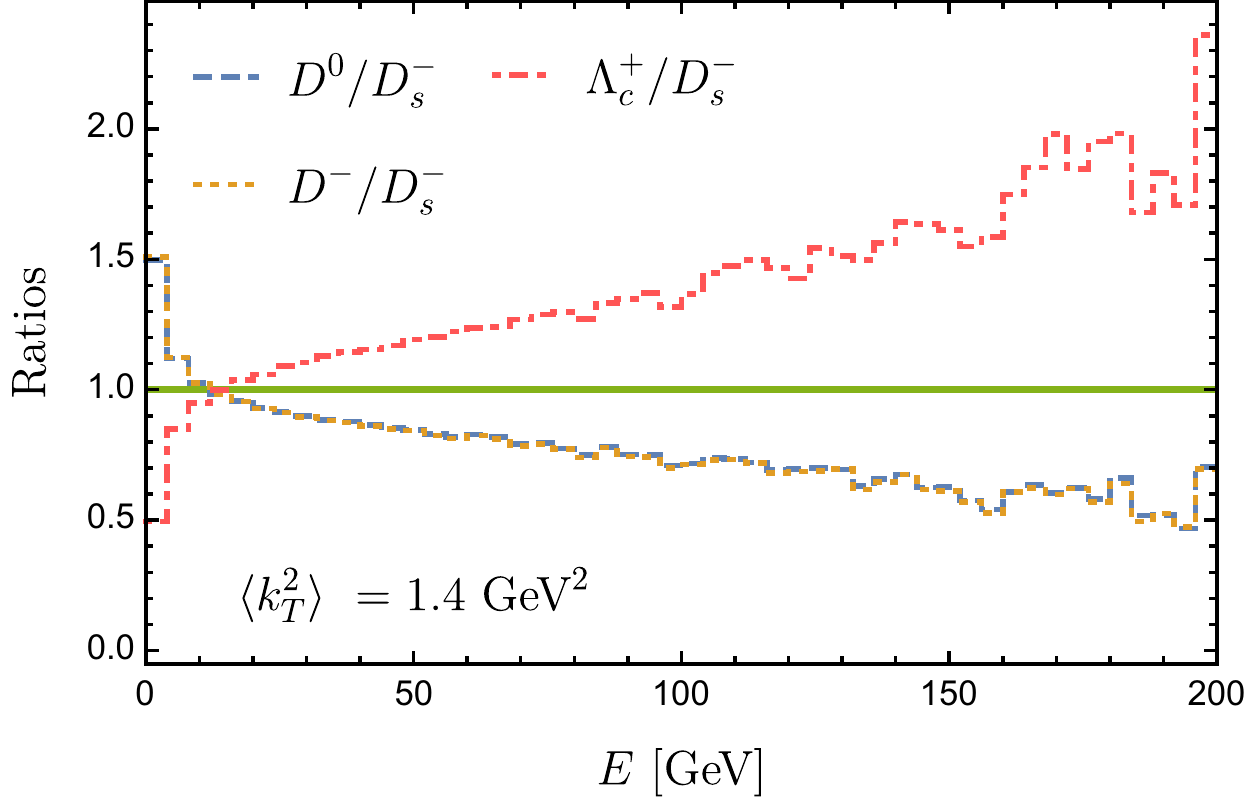}
\caption{The perturbative differential cross section per nucleon for $p$Mo interactions for $E_p=400$ GeV on a fixed target ($\sqrt{s}=27.4$ GeV) to produce a charm hadron,
as a function of the charm hadron
energy, with intrinsic transverse momentum $\langle k_T^2\rangle=1.4$ GeV$^2$ for all charm hadrons (upper figure).
For comparison purposes, the fragmentation fractions have been divided out so the curves all have the same normalizations. 
The ratios of the normalized distributions for $D^0$, $D^-$ and $\Lambda_c^+$ to the $D_s^-$ distribution {are shown in the lower figure}. {The $D^0$ and $D^-$
histograms overlap.}}
\label{fig:dsdxE}
\end{center}
\end{figure}

Fig. \ref{fig:dsdxE} (upper) shows the differential cross section as a function of the charm hadron energy for each of the charm hadrons considered here. 
To compare the distributions, the distributions have been normalized by $1/B_{c\to H}$. Fig. \ref{fig:dsdxE} (lower) shows the ratios of the
normalized distributions relative to the $D_s^-$ distribution as a function of charm hadron energy. {The $D^0$ and $D^-$ histograms overlap.}

\begin{figure}
\begin{center}
\hspace{-0.85cm}
\includegraphics[width=0.75\textwidth]{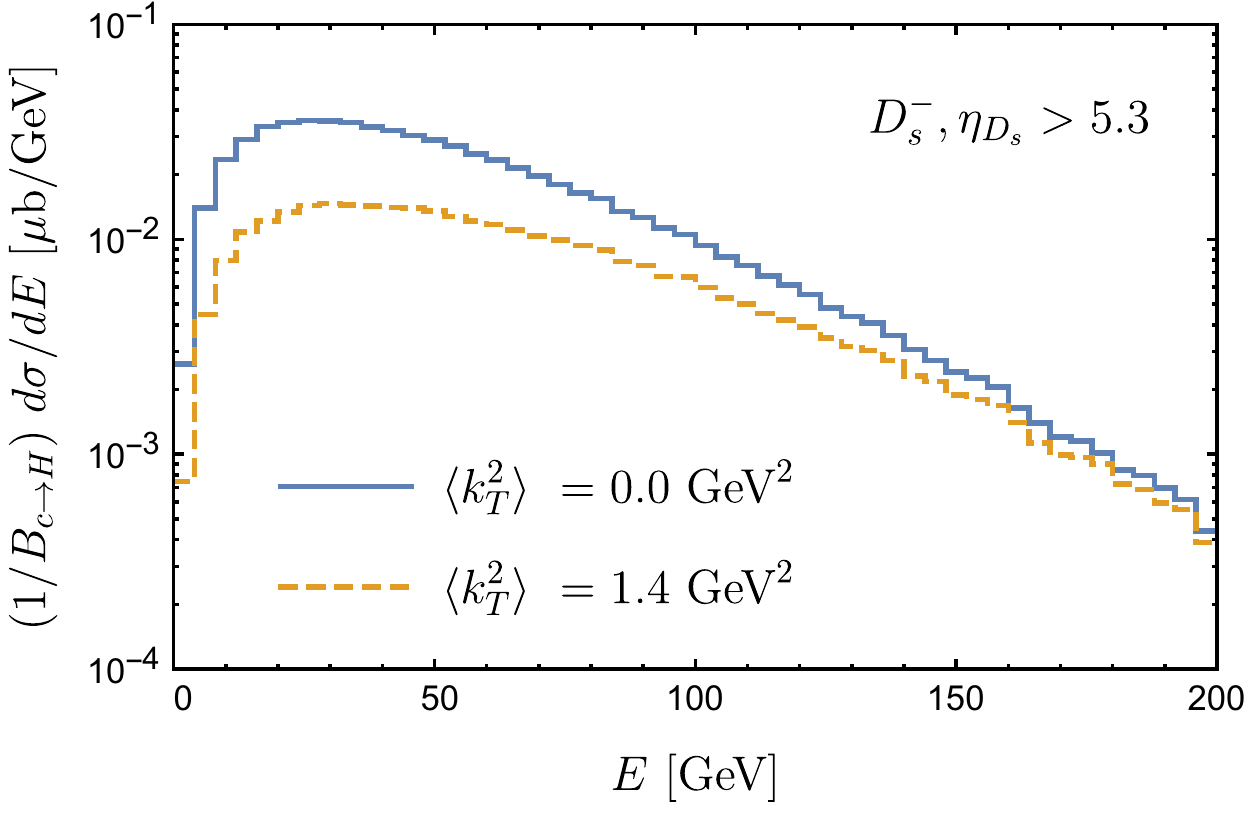}
\caption{The differential cross section per nucleon for $p$Mo interactions {scaled by fragmentation fraction} for $E_p=400$ GeV
and $\langle k_T^2\rangle=0$ (solid) and $1.4$ GeV$^2$ (dashed), as a function of 
charm hadron energy $E$, for perturbative production of $D_s$ with $\eta_{D_s}>5.3$ (dashed line). }
\label{fig:dsdxE-5p3}
\end{center}
\end{figure}

The {charm hadron} energy distributions integrated over the full phase space do not change significantly with the introduction of intrinsic transverse momentum, however, 
the energy distributions of the charm hadrons do change when an angular cut is applied. To illustrate, we show in fig. \ref{fig:dsdxE-5p3}
the energy distribution of the $D_s$ when $\eta_{D_s}>5.3$. {For $\eta_{D_s}>5.3$, the cross section for $D_s$ production including
intrinsic transverse momentum is 7\% of the
total $D_s$ cross section. The intrinsic $k_T$ has a significant impact.
Using the pseudorapidity cut $\eta>5.3$ and no intrinsic $k_T$, the number of low energy $D_s$'s is approximately a factor of two higher 
than with intrinsic $k_T$. 
The inclusion of intrinsic $k_T$}  translates to a lower flux of low energy tau neutrinos at the SHiP detector. This is not unexpected. Since $\langle k_T\rangle\simeq 1$ GeV and $\theta\sim \langle k_T\rangle/E$,  the requirement $\theta\lesssim 0.01$ rad means the energies out to $\sim 100$ GeV are impacted.

\begin{figure}
\begin{center}
\hspace{-0.85cm}
\includegraphics[width=0.75\textwidth]{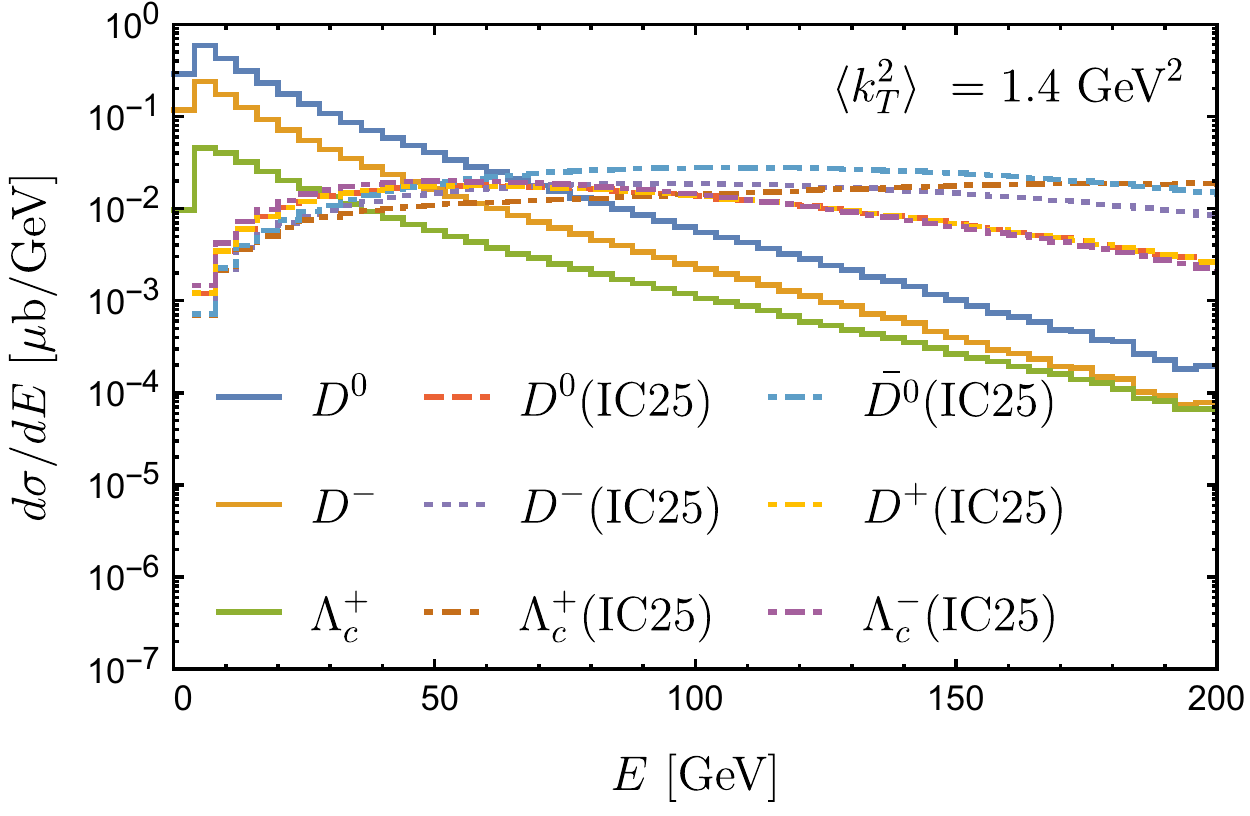}
\includegraphics[width=0.75\textwidth]{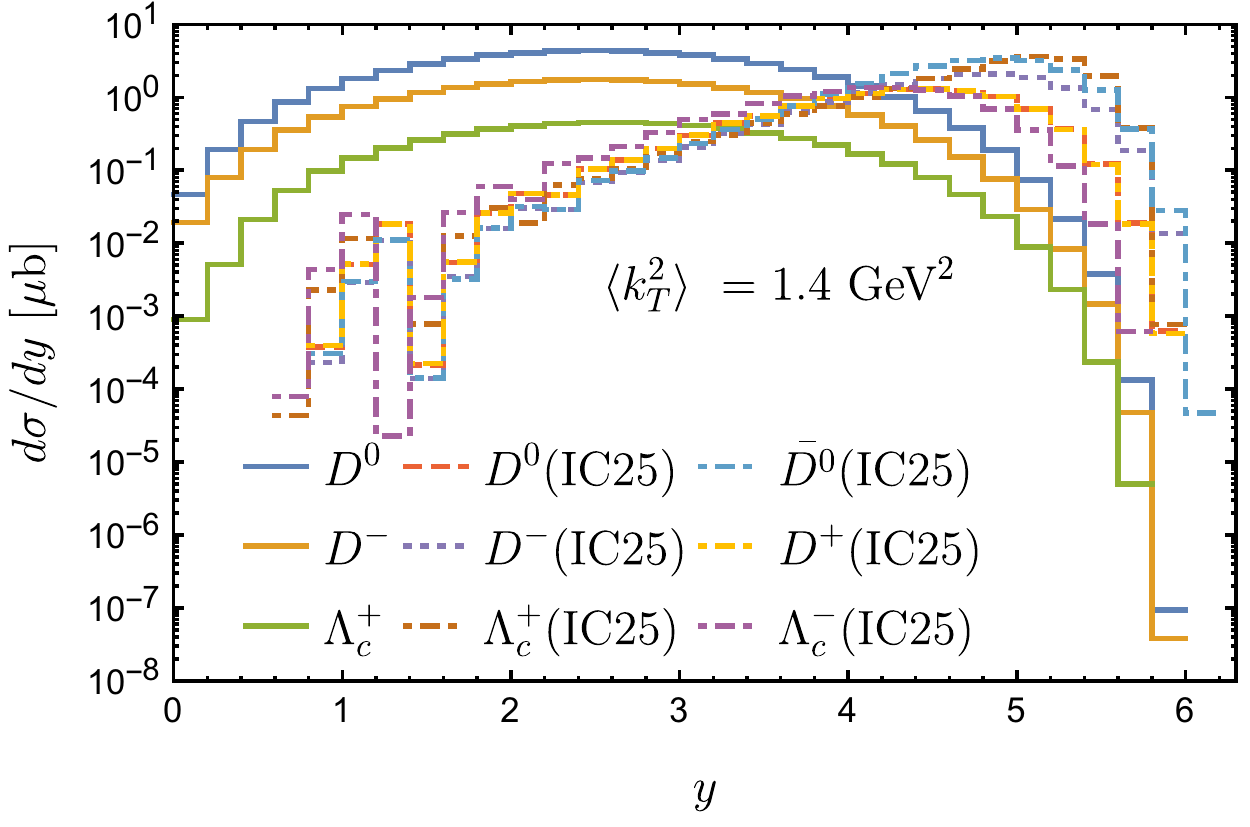}
\caption{(Upper figure) The differential cross section per nucleon for $p$Mo interactions for $E_p=400$ GeV
and $\langle k_T^2\rangle=1.4$ GeV$^2$, as a function of 
charm hadron energy $E$ for perturbative charm production (solid histograms) and intrinsic charm (dashed and dot-dashed histograms). Intrinsic charm is normalized according to ref. \cite{Laha:2016dri} as shown in eq. (\ref{eq:icnorm}){ (IC25)}. (Lower figure) The differential cross section per nucleon for $p$Mo interactions and $\langle k_T^2\rangle=1.4$ GeV$^2$, as a function
of the charm hadron rapidity $y$ in the lab frame for perturbative charm (solid histograms) and intrinsic charm
(IC25) (dashed
and dot-dashed histograms).}
\label{fig:dsdxE-ic}
\end{center}
\end{figure}

The charm hadron distributions for both perturbative production and via intrinsic charm are shown in fig. \ref{fig:dsdxE-ic}
as a function of energy (upper figure) and lab frame rapidity (lower figure). The upper figure
shows that intrinsic charm can dominate the energy distribution of charm hadrons
for hadron energies larger than $E\sim 50-100$ GeV, depending on the normalization of the intrinsic charm contribution.
{Fig. \ref{fig:dsdxE-ic} shows distributions from intrinsic charm with the IC25 normalization of eq. (\ref{eq:icnorm}). } 
{The lower figure shows that} intrinsic charm clearly dominates at high rapidity, as anticipated.
{When a pseudorapidity cut of $\eta>5.3$ is applied to the $D_s$ distribution from intrinsic charm, the cross section is 40\% of the cross section for 
all of phase space.}

\begin{figure}
\begin{center}
\hspace{-0.85cm}
\includegraphics[width=0.75\textwidth]{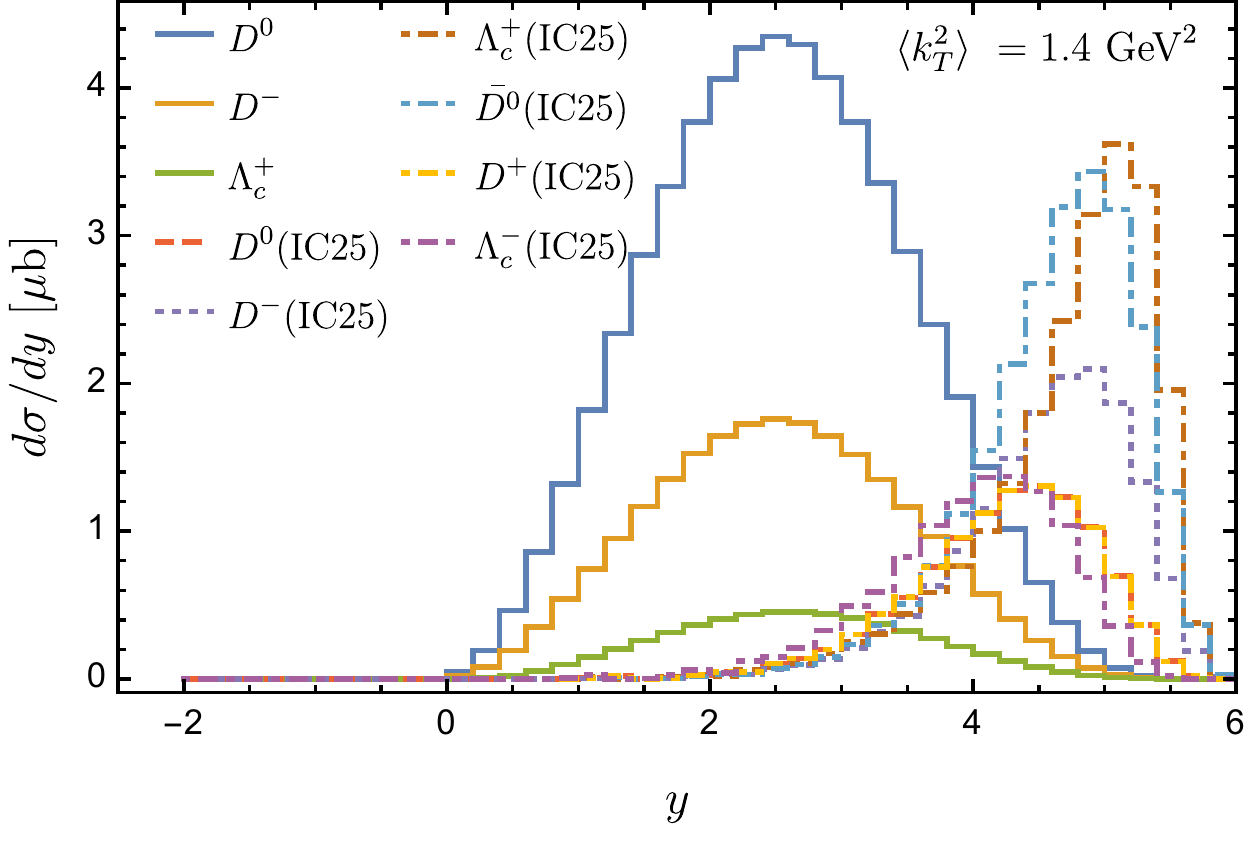}
\caption{The differential cross section per nucleon for $p$Mo interactions and $\langle k_T^2\rangle=1.4$ GeV$^2$, as a function
of the charm hadron rapidity $y$ in the lab frame for perturbative charm (solid histograms) and intrinsic charm
(IC25), normalized
as in fig. \ref{fig:dsdxE-ic} (dashed
and dot-dashed histograms), plotted on a linear scale.
}
\label{fig:dsdy-ic}
\end{center}
\end{figure}

The perturbative result for $D^+$ equals the result for $D^-$, and similarly for the other hadrons, however, the intrinsic charm contributions are different for hadrons incorporating
a charm quark compared to charm antiquark. The $\Lambda_c^+$ are produced at the highest rapidities, as can be  seen in fig. \ref{fig:dsdy-ic}, where perturbative and intrinsic contributions to charm hadron 
rapidities are shown on a linear scale.

\begin{figure}
\begin{center}
\hspace{-0.85cm}
\includegraphics[width=0.7\textwidth]{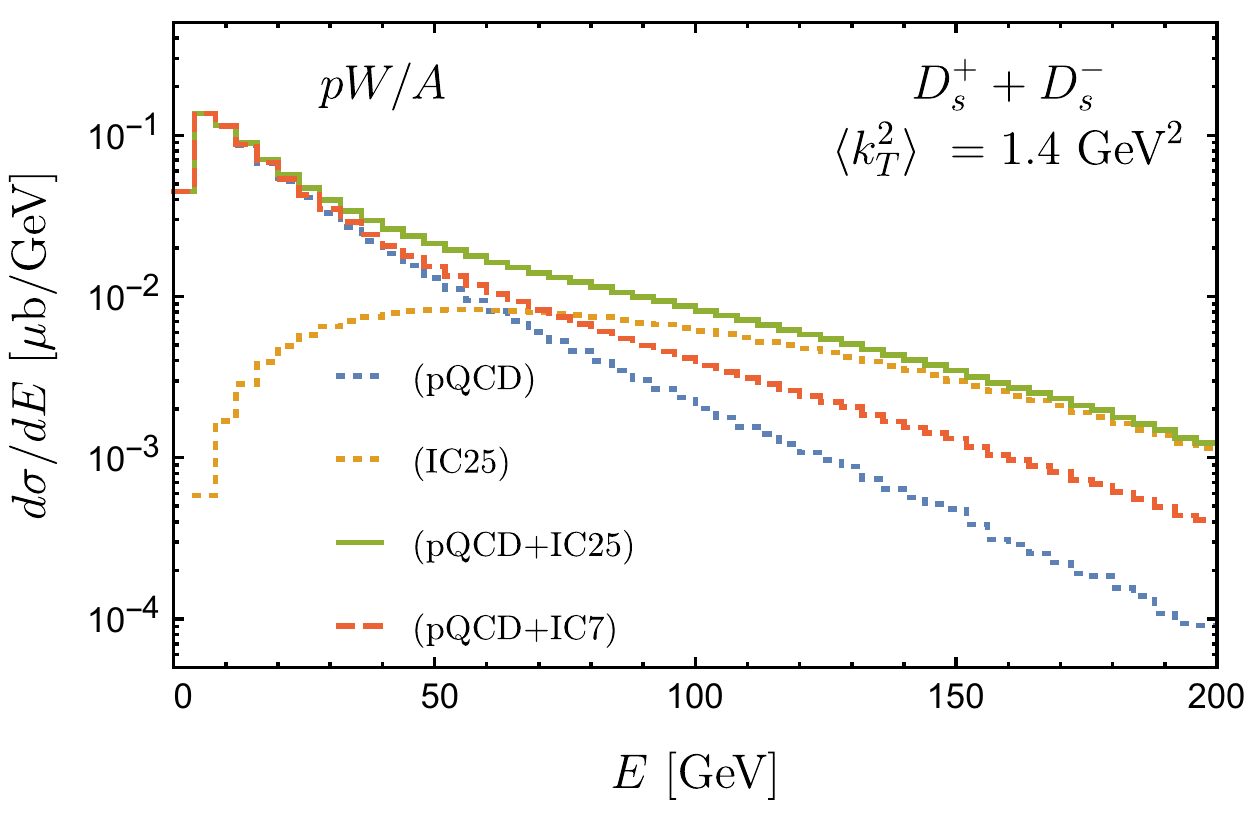}
\caption{The differential cross section per nucleon for $pW$ interactions for $E_p=400$ GeV
and $\langle k_T^2\rangle=1.4$ GeV$^2$, as a function of  $D_s^\pm$
energy $E$ for perturbative $D_s^++D_s^-$ production (blue dotted) and intrinsic charm (orange dotted), where intrinsic charm is normalized according to 
eq. (\ref{eq:icnorm}) (IC25), adjusted to account for a tungsten target. The sum is shown with the solid histogram. The red dashed histogram shows the sum when the intrinsic charm is scaled by 7/25 (IC7).}
\label{fig:dsdxE-icds}
\end{center}
\end{figure}

Perturbative charm and intrinsic charm production of $D_s$'s may be measured in the near future.
The DsTau experiment proposes to study $D_s$ production from a double kink signature, where the decay $D_s\to \tau \nu_\tau$ 
and the $\tau \to \nu_\tau X$ will be observed in an
emulsion detector. They aim for $2.3\times 10^8$ proton interactions in a thin tungsten target to get an estimated 1000 double kink
events when a 20\% detector efficiency is included.
Our evaluation of the differential distribution for $D_s^\pm$
for the tungsten target is shown in fig. \ref{fig:dsdxE-icds}, where we have used the tungsten PDFs from nCTEQ15 \cite{Kovarik:2015cma} in the evaluation. 
The lower blue dotted histogram comes from the
perturbative QCD evaluation with intrinsic transverse momentum $\langle k_T^2\rangle=1.4$ GeV$^2$, while the orange dotted histogram
shows the intrinsic charm contribution to $D_s^\pm$ normalized by eq. (\ref{eq:icnorm}) (IC25). The sum of the two is shown with the
solid green histogram. The red dashed histogram shows the sum with the intrinsic charm reduced by a factor of 7/25, labeled by IC7.
For perturbative QCD production, with a branching fraction of $B(D_s\to \tau \nu_\tau)=5.5\%$ and
the factor of 20\% for detector efficiency,
 the number of double kink events is {881} with our evaluation, integrated over all energy, with {296} events coming
from $D_s^\pm$ with $E>24$ GeV. For the intrinsic charm contribution with the full energy range, the number of events from IC is {277}, 
with {261} events having $D_s^\pm$ energies above 24 GeV. For $E>100$ GeV, there are {19} perturbative events and {95} IC events of this type. The lower normalization factor of 7/25 for IC reduces the number of IC events by a factor of 0.28.
Approximating $x_F\simeq x_E=E_{D_s}/E_p$, {for a power law of the form}
\begin{equation}
\frac{d\sigma}{dx_E}\sim (1-x_E)^{n}\ ,
\end{equation}
{the power is} $n\simeq  10.0$ at large $x_E$ for the perturbative result and
$n\simeq 5.0$ for the sum of perturbative and intrinsic charm as evaluated by this model.
The DsTau experiment should be able to put strong constraints on
the intrinsic charm model of ref. \cite{Gutierrez:1998bc}.

\section{Neutrinos and antineutrinos at SHiP}

{In this section we show energy distributions of neutrino and antineutrinos from charm hadron decays. In sec. 3.1, the NLO perturbative
QCD contributions are evaluated first, followed by the inclusion of the much more uncertain component from intrinsic charm.
The neutrino energy distributions from charm hadron decays are converted to the number of events per GeV 
at SHiP for muon and tau neutrinos and antineutrinos. We represent the SHiP detector as a column depth $L=66.7$ g/cm$^2$ of lead and
convert the neutrino energy distributions to events using the neutrino charged current cross section. These results are shown in sec. 3.2.
In sec. 3.3, we describe an approximation for the number of $\nu_\mu$ and $\bar{\nu}_\mu$ from
pion and kaon decays (with more details in appendix C). We evaluate the muon charge asymmetry from muon neutrino and antineutrino
charged current interactions, where the muon neutrino and antineutrino fluxes come from pion, kaon and charm hadron decays, 
where the charm contribution is computed both with and without intrinsic charm.} 

\subsection{Neutrino and antineutrino energy distributions from prompt decays of charm}

\begin{figure}
\begin{center}
\hspace{-0.85cm}
\includegraphics[width=0.75\textwidth]{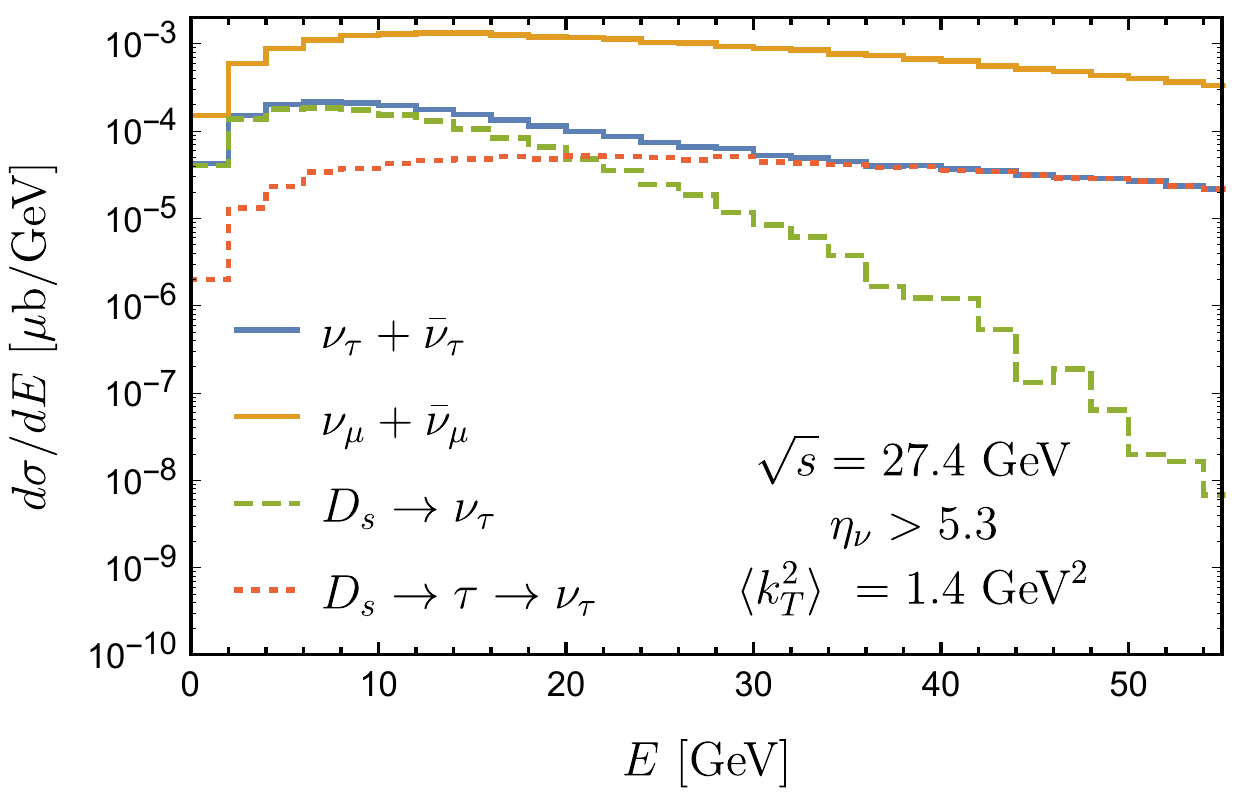}
\caption{The perturbative differential cross section per nucleon for $p$Mo interactions for $E_p=400$ GeV 
and $\langle k_T^2\rangle=1.4$ GeV,$^2$ as a function of 
neutrino energy for $\nu_\mu+\bar{\nu}_\mu$ (upper solid histogram) and
$\nu_\tau+\bar{\nu}_\tau$ (lower solid histogram) for  $\eta_\nu>5.3$. The separate contributions
to $\nu_\tau+\bar{\nu}_\tau$ from the direct decay ($D_s\to \nu_\tau$) and
chain decay ($D_s\to \tau\to \nu_\tau$) are shown with long and short dashed histograms, respectively.}
\label{fig:dsdE-pertnu}
\end{center}
\end{figure}

{The neutrinos from prompt decays of charm hadrons at the SHiP detector, including the full three-dimensional kinematics, are shown in this section.
The $D_s$ decay $D_s\to \tau \nu_\tau$, with the $\tau\to \nu_\tau X$ that depends on the tau polarization, is an interesting
feature. We refer to the tau neutrino produced directly in the $D_s\to \tau \nu_\tau$ as the ``direct neutrino'' and
the neutrino that comes from the tau decay as the  ``chain decay neutrino.''  For reference, we include in appendix B
some details
of our numerical implementation of the production of direct and chain decay tau neutrinos from $D_s$ decays.

The NLO QCD results for the neutrino differential cross section as a function of neutrino energy are shown in fig. \ref{fig:dsdE-pertnu}
for $E_p=400$ GeV. The distributions are the sum of neutrinos and antineutrinos. } The neutrino
or antineutrino rapidity is required to be  $\eta_\nu>5.3$. The upper curve shows the charm contribution to the energy distribution
of $\nu_\mu$ and $\bar{\nu}_\mu$. The lower solid curve shows the tau neutrino plus antineutrino distribution, the sum of the dashed (direct decays) and dotted (chain decays from tau) curves. The direct neutrino has a much smaller energy than the tau in the $D_s$ rest frame. {In the lab frame, even though the neutrino from  the tau decay $D_s\to \tau\to \nu_\tau$  has a fraction of the tau energy, itself a fraction of the $D_s$ energy,  the chain decay neutrino  has more energy on average than the direct neutrino.}  The cross-over from direct decay dominated to chain decay dominated is at $E_\nu\sim 20$ GeV. The prompt muon neutrino plus antineutrino differential cross section is approximately a factor of ten larger than that of prompt tau neutrinos plus antineutrinos. Since charm is produced with anticharm, the  differential cross sections to produce $\nu_\ell$ and $\bar{\nu}_\ell$ are the same
for a given $\ell$.
We approximate the electron neutrino and muon neutrino differential cross sections to be equal since the charged lepton masses $m_e$ and $m_\mu$ are small compared to the charm meson masses. 

\begin{figure}
\begin{center}
\hspace{-0.85cm}
\includegraphics[width=0.7\textwidth]{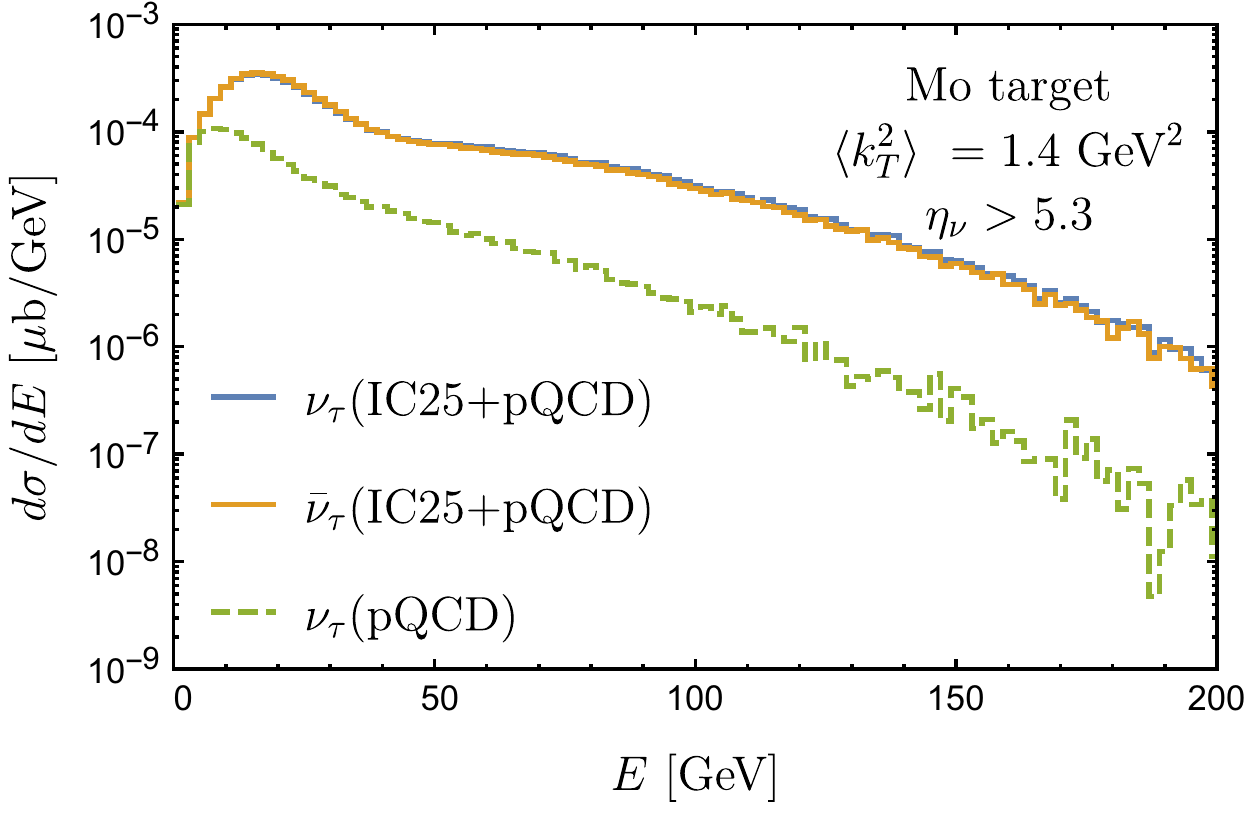}
\caption{The perturbative differential cross section per nucleon (dashed)  $(d\sigma(pA)/dE)/A$  for $E_p=400$ GeV, $A=96$
and $\langle k_T^2\rangle=1.4$ GeV,$^2$ as a function of 
neutrino energy for perturbative production of $\nu_\tau$ (lower dashed histogram) and from the
sum of perturbative and intrinsic charm on a molybdenum target for $\nu_\tau$ and $\bar{\nu}_\tau$
(upper solid histograms), for  $\eta_\nu>5.3$. The intrinsic charm contributions are normalized with eq. (\ref{eq:icnorm}) labeled IC25.} 
\label{fig:dsdE-p+ic-nutau}
\end{center}
\end{figure}

It is not surprising that with the normalization of intrinsic charm discussed in the previous section, intrinsic charm contributions
to the neutrino differential cross section in the very forward region overwhelm the perturbative neutrino differential cross section,
as shown in fig. \ref{fig:dsdE-p+ic-nutau} for the tau neutrino and antineutrino distributions for protons incident on a molybdenum target at SHiP. 
The lower dashed histogram shows the perturbative tau neutrino energy distribution (equal to the perturbative tau antineutrino energy distribution),
while the solid lines show the sum of intrinsic and perturbative tau neutrino, and separately, tau antineutrino energy distributions.
{The intrinsic charm  is normalized with eq. (\ref{eq:icnorm}), IC25,} where with $A=96$, the {nuclear target} normalization factor is $A^{0.71}/A= 0.27$. The default
intrinsic transverse momentum $\langle k_T^2\rangle = 1.4$ GeV$^2$ is used, with $\eta_\nu>5.3$.

\begin{figure}
\begin{center}
\hspace{-0.85cm}
\includegraphics[width=0.7\textwidth]{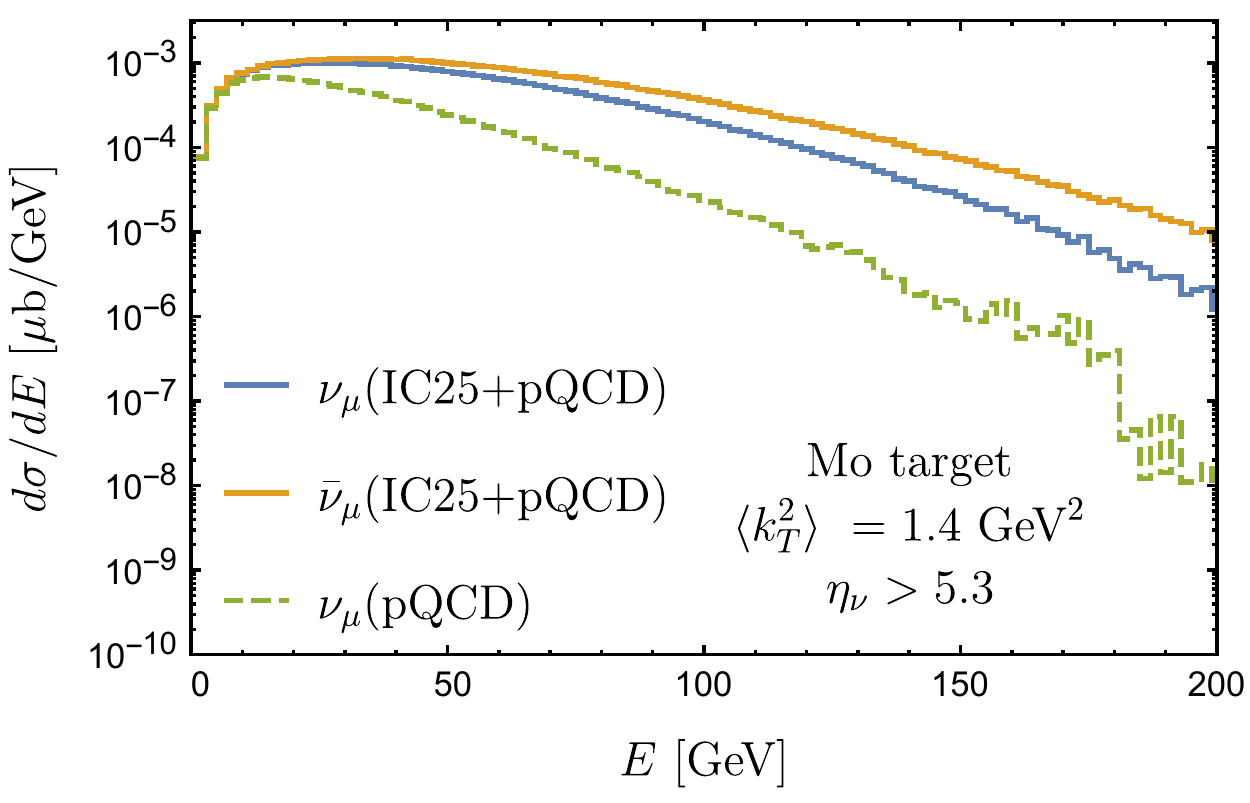}
\caption{
The differential cross section per nucleon (dashed)  $(d\sigma(pA)/dE)/A$  for $E_p=400$ GeV, $A=96$
and $\langle k_T^2\rangle=1.4$ GeV,$^2$ as a function of 
neutrino energy for perturbative production of $\nu_\mu$ (lower dashed histogram) and from the
sum of perturbative and intrinsic charm on a molybdenum target for $\nu_\mu$ and $\bar{\nu}_\mu$
(upper solid histograms), for  $\eta_\nu>5.3$. The intrinsic charm contributions are normalized with eq. (\ref{eq:icnorm}) labeled IC25.}
\label{fig:dsdE-intrinsicnu}
\end{center}
\end{figure}

The dashed line shown in fig. \ref{fig:dsdE-intrinsicnu} shows the $\nu_\mu$ energy distribution $d\sigma/dE$ per nucleon in the target, equal to the $\bar{\nu}_\mu$
energy distribution evaluated using NLO perturbative QCD of $c\bar{c}$ with $\langle k_T^2\rangle=1.4$ GeV$^2$
for $\eta_\nu>5.3$. The solid lines show
the sum of perturbative charm and intrinsic charm contributions to the muon neutrino (lower solid histogram) and muon antineutrino
(upper solid histogram) energy
distributions for the same $\langle k_T^2\rangle$ and rapidity restriction. 
Since charm hadrons $D^-$ and $\bar{D}^0$ are favored over their charged
conjugates with this model for intrinsic charm momentum distributions, the production rate for {muon} anti-neutrinos
is larger than for {muon} neutrinos. Above $E\sim 20$ GeV, the asymmetry between particle
and antiparticle perturbative production {is modified by intrinsic charm. It} will be apparent if the intrinsic
charm contribution is normalized by IC25 (eq. (\ref{eq:icnorm}).) We discuss the muon neutrino-antineutrino asymmetry below, where we show that the enhanced flux of muon antineutrinos compensates in part for the lower antineutrino cross section in the detector.

Intrinsic charm dominates muon neutrino and antineutrino energy
distributions for most of the energy range. The intrinsic charm cross section normalization can be constrained at SHiP from high energy muon neutrino and antineutrino measurements. Indeed, SHiP is better positioned to constrain the normalization of the intrinsic charm contribution than current LHC experiments. At LHC center-of-mass
energies, the intrinsic charm cross section is a much smaller fraction of the perturbative charm cross section. 

\begin{figure}
\begin{center}
\hspace{-0.85cm}
\includegraphics[width=0.7\textwidth]{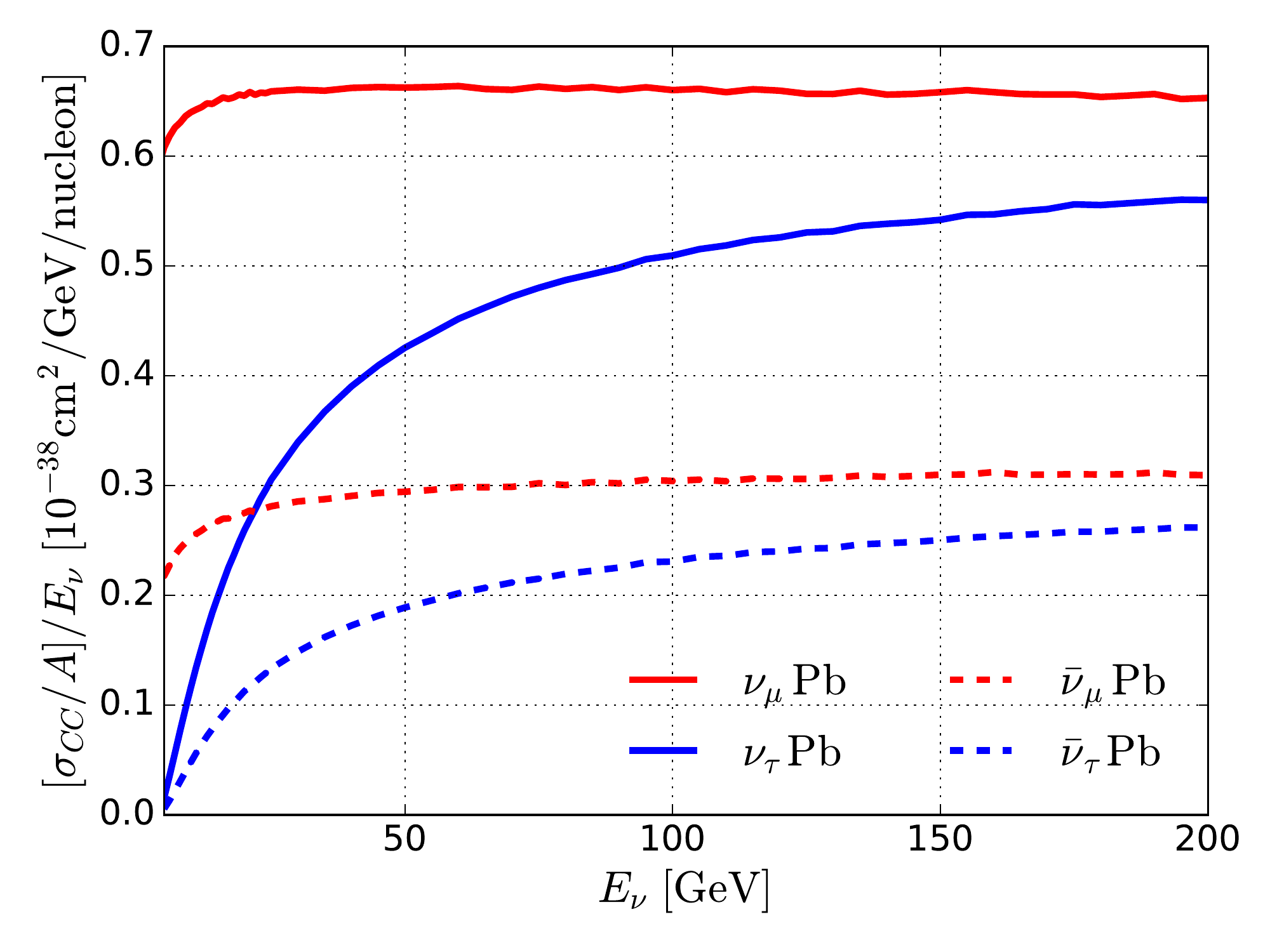}
\caption{The deep-inelastic neutrino and antineutrino charged current cross section per nucleon scaled by (anti-)neutrino energy, for muon and tau (anti-)neutrinos scattering on lead, as a function of neutrino energy. We require that the hadronic final state energy be larger than
$W_{\rm min}=m_p+m_\pi$.}
\label{fig:sigmanu}
\end{center}
\end{figure}

\begin{table}[tb]
\begin{center}
\begin{tabular}{|l|c|c|c|c| }
\hline
$E$ [GeV] & $\sigma (\nu_\mu)/E$ &$\sigma (\bar{\nu}_\mu)/E$&$\sigma (\nu_\tau)/E $&$\sigma (\bar{\nu}_\tau)/E$\\
\hline
1 & 0.360 &  $9.83\times 10^{-2}$&- & - \\
\hline
2 & 0.514 & 0. 151 &- & - \\
\hline
3 & 0.565 & 0.181&- & - \\
\hline
4 & 0.592  & 0.203&- & - \\
\hline
5 & 0.608  & 0.218& $1.58\times 10^{-2}$ &  $5.55\times 10^{-3}$\\
\hline
6 & 0.618 & 0.227& $3.53\times 10^{-2}$ & $1.32\times 10^{-2}$\\
\hline
7 & 0.626 & 0.237 & $5.60\times 10^{-2}$ & $2.18\times 10^{-2}$\\
\hline
8 & 0.631 & 0.243 & $7.69\times 10^{-2}$  & $3.06\times 10^{-2}$\\
\hline
9 & 0.637 & 0.248 & $9.70\times 10^{-2}$  & $3.95\times 10^{-2}$\\
\hline
10  & 0.640 & 0.252 & $0.116$ & $4.81\times 10^{-2}$\\
\hline
20 & 0.655 & 0.275& 0.259 & 0.113\\
\hline
30 & 0.661 & 0.286 & 0.340 & 0.149\\ 
\hline
50 & 0.663 &  0.294& 0.426 &  0.189\\ 
\hline
100 & 0.660 & 0.304& 0.510 & 0.231\\
\hline
200 & 0.653 & 0.309& 0.561 & 0.262\\
\hline
300 & 0.648 & 0.314 & 0.580 & 0.27\\
\hline
400 & 0.643 & 0.315 & 0.586 & 0.282\\
\hline
\end{tabular}
\caption{\label{table:numutaucc} 
The deep-inelastic charged current cross section divided by energy for $\nu_\mu$, $\bar{\nu}_\mu$, $\nu_\tau$ and
$\bar{\nu}_\tau$ interactions with lead, per nucleon,
evaluated using the Pb PDFs of nCTEQ15 \cite{Dulat:2015mca}, in units of $10^{-38}\ {\rm cm}^2/{\rm GeV}$, with hadronic final state energy $W>m_p+m_\pi$.}
\end{center}
\vspace{-0.6cm}
\end{table}

\subsection{SHiP neutrino events from prompt decays of charm}

{The conversion of the differential neutrino energy distribution to events with a charged lepton requires the neutrino charged current cross section. The SHiP detector will have thin emulsion films interleaved with lead bricks \cite{Anelli:2015pba}. 
For the flux of neutrinos from }
charm decays, we write
\begin{equation}
\phi_\nu = \frac{dN_\nu}{dE} = \frac{N_p}{\sigma_{pA}} \frac{d\sigma(pA\to \nu  X)}{ dE}\ .
\end{equation}
As in ref. \cite{Alekhin:2015byh}, we take $N_p=2\times 10^{20}$ protons 
and {hadronic cross section per nucleon}
$\sigma_{pA}/A=\sigma_{pN}=10.7$ mb. The {number of events per GeV} is
\begin{equation}
\phi_{\rm evts} = \phi_\nu \frac{L}{\lambda_\nu}
\end{equation}
where $\lambda_\nu=(N_A \sigma(\nu A)/A)^{-1}$ and $L = 66.7$ g/cm$^2$ for lead in the detector.

{To evaluate the neutrino cross sections with lead relevant for SHiP, we use the nCTEQ15 parton distribution functions for lead \cite{Dulat:2015mca}. 
For tau neutrino interactions, we include the $m_\tau/E_\nu$ corrections to the differential cross section
\cite{Albright:1974ts,Kretzer:2002fr,Kretzer:2003iu,Jeong:2010nt}. 
We use the low momentum transfer extrapolation of the deep-inelastic scattering (DIS) structure functions
outlined in refs. \cite{Reno:2006hj,Jeong:2010nt}.
Shown in fig. \ref{fig:sigmanu} are the neutrino
and antineutrino charged-current 
cross sections per nucleon per neutrino energy as a function of incident neutrino energy for the production of muons and taus. 
We require that the hadronic final state energy $W$ be larger than $W_{\rm min}=m_p+ m_\pi$  for DIS.
At low energy for $\nu_\mu$ interactions, the impact of $W>W_{\rm min}$ can be seen in fig. \ref{fig:sigmanu} for muon neutrinos and antineutrinos.
For tau neutrinos, the suppression of $\sigma/E$ comes from both $W>W_{\rm min}$ and 
tau mass kinematic corrections. The tau mass corrections persist out to relatively high energies. The muon neutrino charged current
cross section is about a factor of 1.5 larger than the tau neutrino cross section at $E_\nu=50$ GeV. 
For reference, table \ref{table:numutaucc} lists the charged current cross sections for scattering with a lead target by muon neutrinos and antineutrinos, and tau neutrinos and antineutrinos 
{for energies up to $E_\nu=400$ GeV. This is beyond the energy range of neutrinos at SHiP, but it is potentially of interest for $E_p=
800$ GeV protons incident on a beam dump. 
The muon neutrino cross section is still 10\% larger than the tau neutrino cross section for $E_\nu=400$ GeV, even though
$E_\nu\gg m_\tau$.  

The uncertainties in the cross sections for neutrino and antineutrinos incident on lead can be approximated by considering the PDF uncertainties as quantified by the 32 variants to the nCTEQ15 PDFs for lead targets.  The cross section uncertainties for $\nu_\tau$ and
$\bar{\nu}_\tau$ DIS interactions in lead using these PDF variants are smaller than $\pm 3\%$ for the energies considered here, except for the lowest energies, below $\sim 10$ GeV.}
}

\begin{figure}
\begin{center}
\hspace{-0.85cm}
\includegraphics[width=0.75\textwidth]{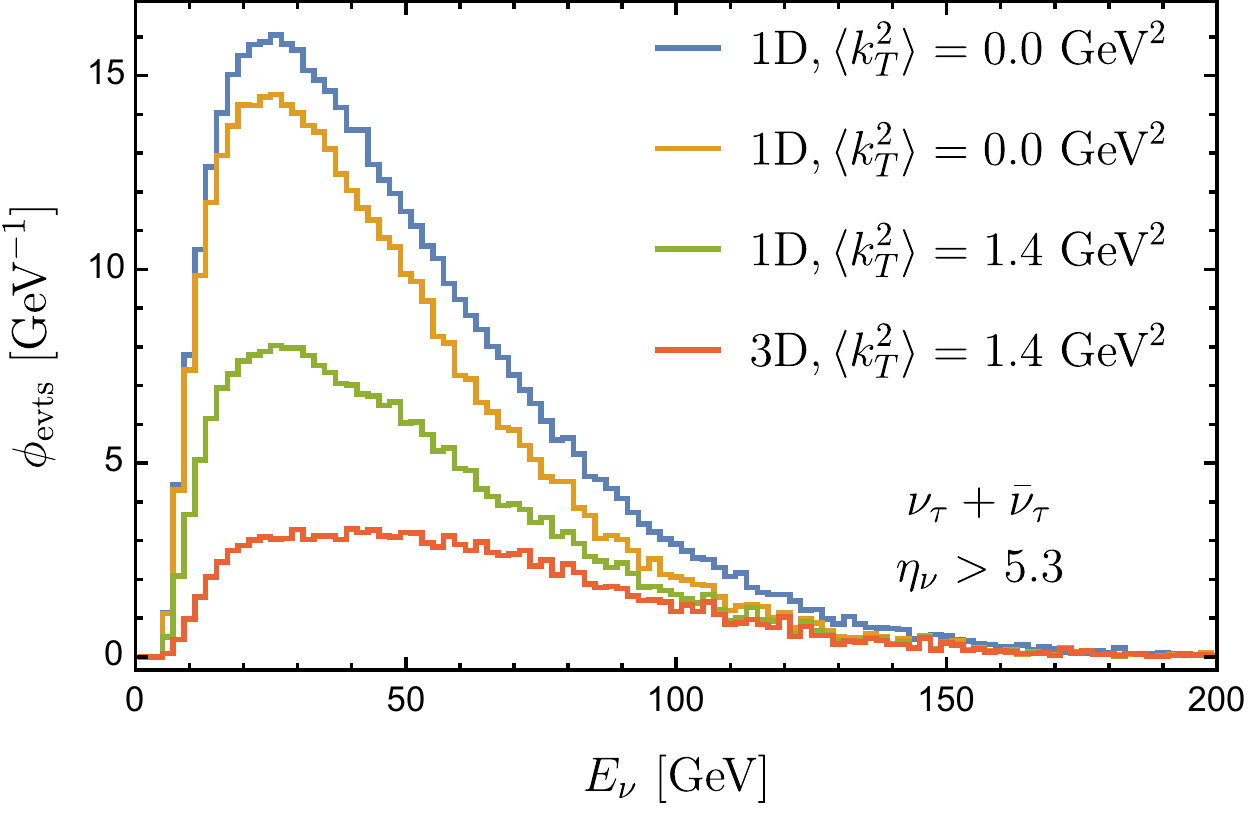}
\caption{From perturbative QCD, the sum of tau neutrino plus tau antineutrino events per GeV for  $\eta_\nu>5.3$ as a function of
incident neutrino energy, for an 400 GeV proton beam with $2\times10^{20}$ protons incident on a molybdenum target,
assuming a detector with column depth $L=66.7$ g/cm$^2$ of lead, for different approximations in evaluating angular
dependence of the neutrino: $\langle k_T^2\rangle=0$ and $\langle k_T^2\rangle = 1.4$ GeV$^2$, with collinear (1D) and full
three-dimensional (3D) treatment of decays. {The upper histogram comes from using the CT14 PDFs for the charm production, while the remaining histograms show results
with nCTEQ15 PDFs.}}
\label{fig:events}
\end{center}
\end{figure}

Fig. \ref{fig:events} shows the impact of intrinsic transverse momentum and the full three-dimensional kinematics for the sum of tau neutrino plus antineutrino events for $\eta_\nu>5.3$, as a function of neutrino energy, evaluated using NLO perturbative QCD.
The upper blue histogram shows the number of events {per GeV} for tau neutrinos plus antineutrinos, evaluated using {most of} the same approximations as in ref. \cite{Alekhin:2015byh}: that the differential cross section for charm quark production, evaluated with {CT14} PDFs with no intrinsic transverse momentum {but with fragmentation, in the approximation that the charm quark momentum is in the same direction relative to the incident proton beam as the neutrino from its decay (the ``collinear approximation'' that $\eta_\nu\simeq \eta_c$). Here we show for $\eta_\nu>5.3$ as an approximate representation of the 
detector solid angle. }
Using CT14 PDFs reduces the number of events by $\sim 10\%$ compared to the number of events with CT10 PDFs
{used in ref. \cite{Alekhin:2015byh}. }

{The three lower histograms in fig. \ref{fig:events} are evaluated using the best fit nCTEQ15 PDFs for the free proton incident on the
fixed molybdenum target.}
With {only the addition of } intrinsic
transverse momentum $\langle k_T^2\rangle=1.4$ GeV$^2$, the {number of events} is 62\% of the {number of events} without intrinsic transverse
momentum. The three dimensional decay kinematics further reduces the overall number of events. The number of events goes from
939 to 272 events for $N_p=2\times 10^{20}$ with these parameter choices for 
NLO perturbative production of charm with fragmentation to $D_s$ mesons that decay to $\tau \nu_\tau$.

\begin{figure}
\begin{center}
\hspace{-0.85cm}
\includegraphics[width=0.7\textwidth]{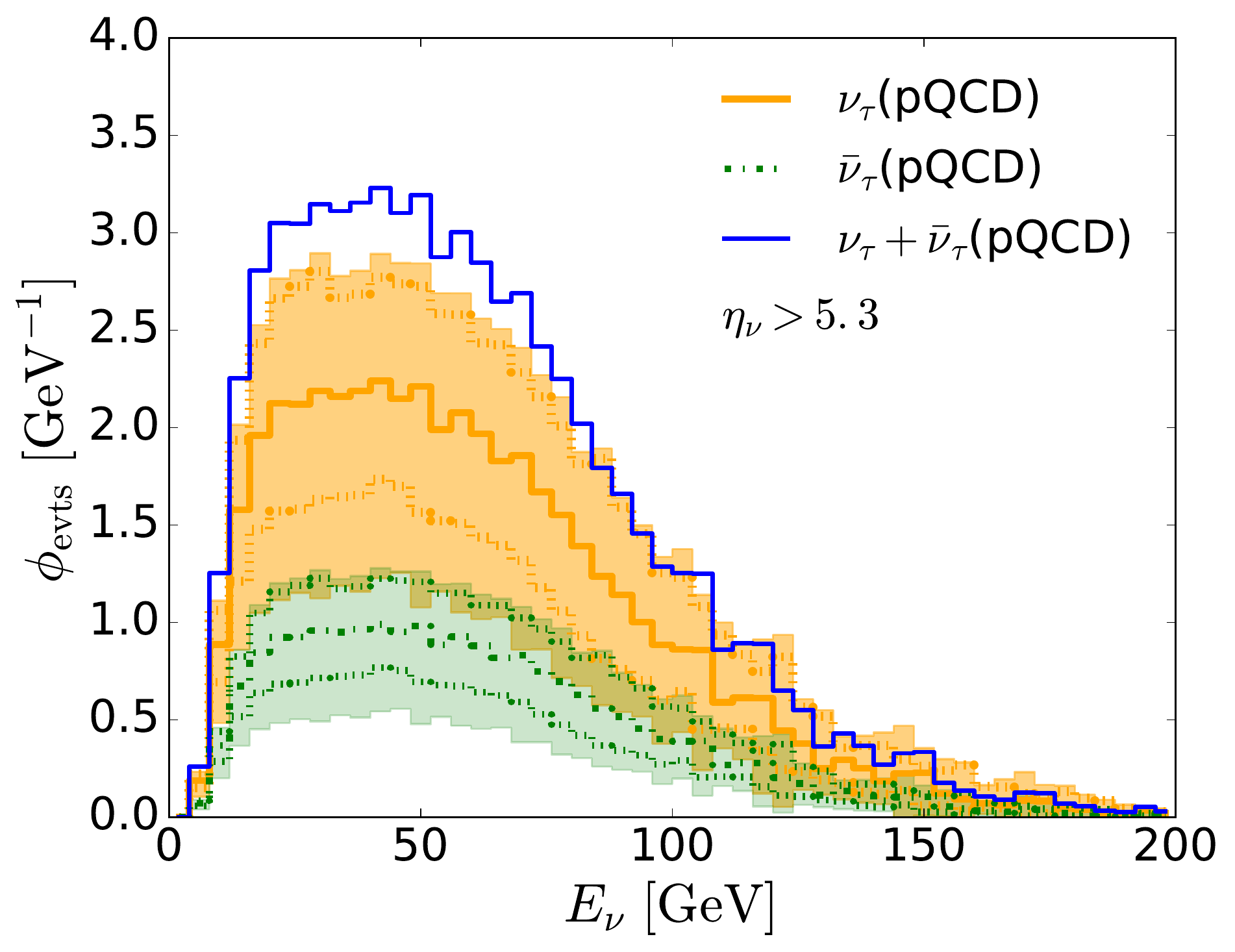}
\includegraphics[width=0.7\textwidth]{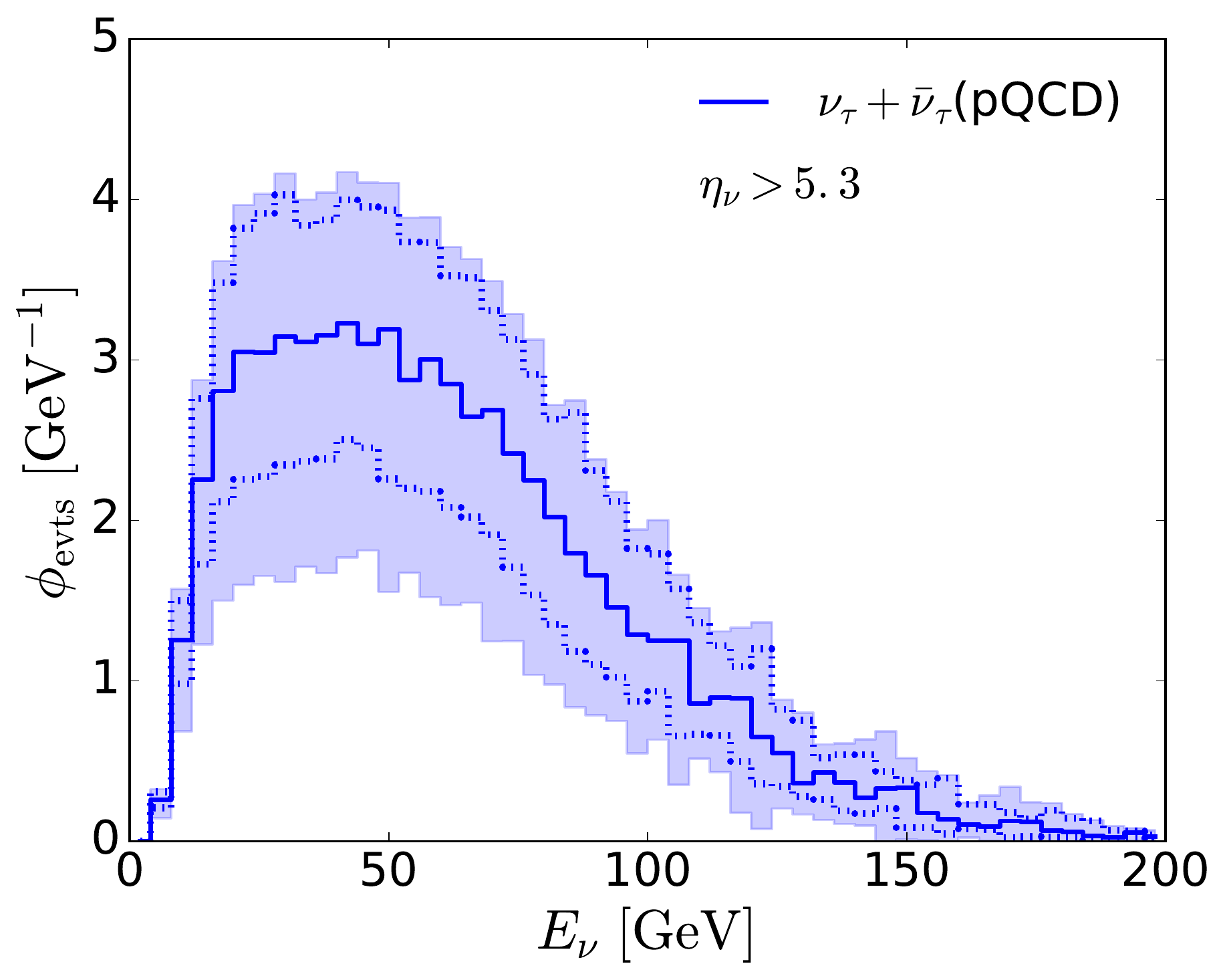}
\caption{From perturbative QCD, the number of events per GeV for tau neutrinos and tau antineutrinos (upper figure)
and the  sum (soild blue histogram in upper figure, and lower figure) for  $\eta_\nu>5.3$ as a function of
incident neutrino energy, for an 400 GeV proton beam with $2\times10^{20}$ protons incident on a molybdenum target,
assuming a detector with column depth $L=66.7$ g/cm$^2$ of lead. The central curve has the scale choices $(\mu_R,\mu_F) = (1.6,2.1)m_c$.
The colored bands bracketed with dotted lines reflect the range of scales from 
$(\mu_R,\mu_F) = (1.48,1.25)m_c$ and $(\mu_R,\mu_F)=(1.71,4.65)m_c$ and intrinsic $\langle k_T^2\rangle=1.1-1.7$ GeV$^2$.  The shaded band  includes additional uncertainties
from the nuclear PDFs, and for the lower band, from using $m_c=1.4$ GeV and $(\mu_R,\mu_F) = (1.6,2.1)m_T$, added in quadrature.}
\label{fig:events-scale}
\end{center}
\end{figure}

{Our main results for tau neutrinos and antineutrinos at SHiP, evaluated with NLO perturbative QCD, are shown in fig. \ref{fig:events-scale}.
In the upper figure, we show separately the tau neutrino and tau antineutrino number of events per GeV. The solid histograms  show the number of events per GeV for the central scale choices with $m_c=1.27$ GeV and $\langle k_T^2\rangle=1.4$ GeV$^2$.
The difference in $\phi_{\rm evts}$ from $\nu_\tau$ and $\bar{\nu}_\tau$ comes from the difference
in the neutrino and antineutrino cross sections. In the lower figure, the solid histogram shows the sum of tau neutrino plus antineutrino events per GeV.} 

{The shaded bands reflect the uncertainties in the evaluation of the number of events at SHiP.}
For a fixed charm quark mass, the choice of the scale factor has the largest impact on the predicted number of events. As in ref. \cite{Alekhin:2015byh}, we use a range of scales guided by {the data constrained analysis of}
ref. \cite{nelson2013narrowing}, namely, $(\mu_R,\mu_F) = (1.48,1.25)m_c$ and $(\mu_R,\mu_F)=(1.71,4.65)m_c$ 
with $m_c=1.27$ GeV to bracket the predicted
charm production cross sections. {The choice of intrinsic transverse momentum $\langle k_T^2\rangle=1.1-1.7$ GeV$^2$ introduces an uncertainty
that is between $\sim 3-6$ times smaller than the effect of the scale uncertainty for $E_\nu<100$ GeV.}
The dotted histograms show the range of predictions adding in quadrature the scale and $\langle k_T^2\rangle$ excursions from the central value.
The range
in the number of events is 195 events to 357 events (sum of tau neutrino plus antineutrino) accounting just for the scale variation using $\mu_{R,F}\sim m_c$ and $\langle k_T^2\rangle=1.1-1.7$ GeV$^2$. The central choices give 272 events.
The number of events vary by less than $\sim \pm 40\%$ below $E=100$ GeV, less than $\sim \pm 30\%$ below 50 GeV.

{Fig. \ref{fig:events-scale} also shows a broader uncertainty band for the separate tau neutrino and antineutrino number of events per GeV (upper plot)
and for the sum (lower plot). The full uncertainty band includes, in addition to the scale dependence and choice of $\langle k_T^2\rangle$, a number of other
uncertainties added in quadrature. 
The 33 sets of nuclear PDFs provided in the nCTEQ15 distribution translate to an uncertainty in the number of tau neutrinos plus antineutrinos
that increase with energy, from a few percent error at low energies, to $\sim 10\%$ for $E_\nu=50$ GeV and $\sim 20\%$ for $E_\nu=100$ GeV.
Using the nCTEQ15 set for krypton rather than silver decreases the distribution as a function of energy with a smaller, but similar correction.
The nCTEQ15 results are about $14\%$ lower than using CT14 PDFs for the free proton, so we add (in quadrature)
an additional 14\%  uncertainty to the upper band.
The uncertainty band also includes decreases in the distribution that come from 
 using $m_c=1.4$ GeV and $\mu_{R,F}\sim m_T$. Each of these two changes to the inputs of the calculation decreases the predicted number
 of event per GeV comparably to choosing the scales to be $(\mu_R,\mu_F)=(1.71,4.65)m_c$ with $m_c=1.27$ GeV.
We remark, however, that neither choice gives a total $c\bar{c}$ cross section within $3\sigma$ of the measured value.
An additional uncertainty included is the small (of order 3\%) uncertainty that comes from evaluating the neutrino and antineutrino lead cross sections with the 32 nCTEQ15 PDF variants. 
 For the total number of events shown in the lower figure, the lower limit of the band amounts to
 134 events.  The number of events using the upper limit of the band is 378.}

\begin{figure}
\begin{center}
\hspace{-0.85cm}
\includegraphics[width=0.7\textwidth]{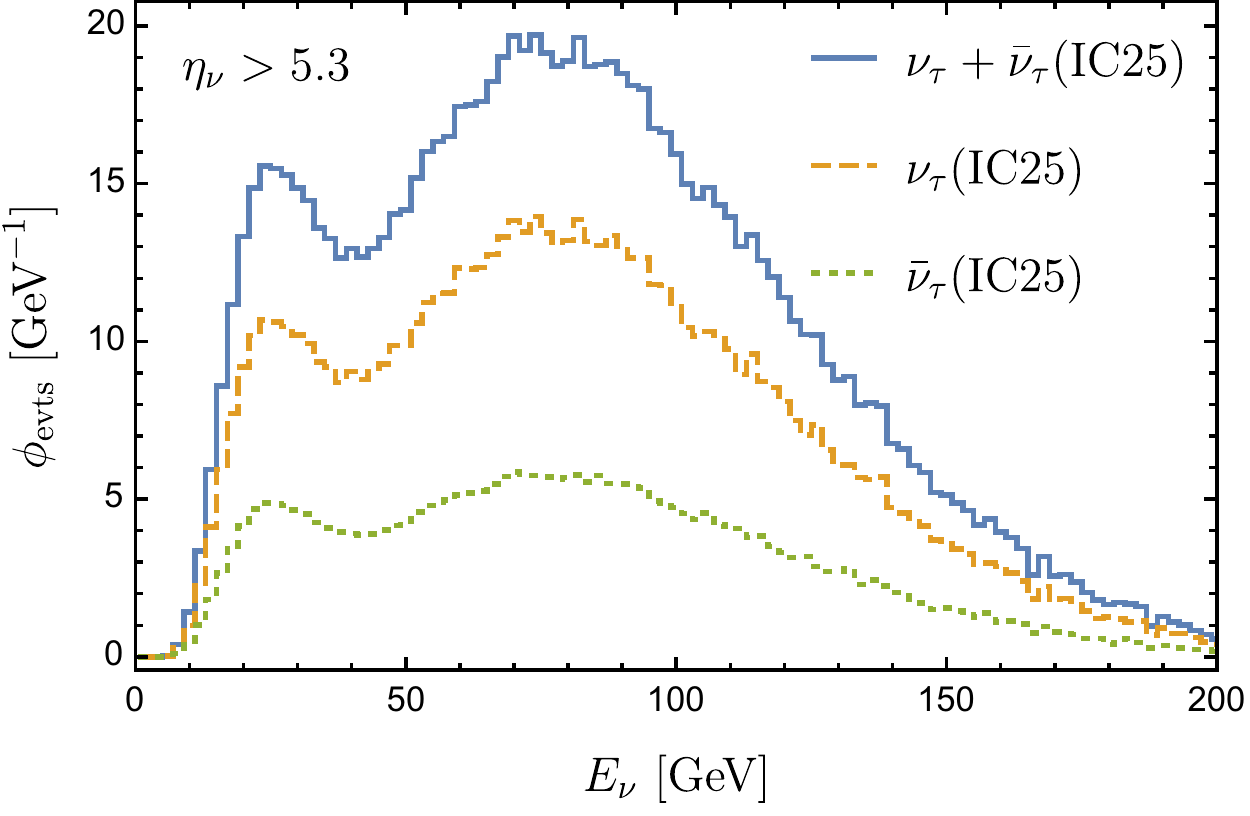}
\caption{The separate tau neutrino, tau antineutrino and sum of tau neutrino plus tau antineutrino events per GeV for  $\eta_\nu>5.3$ as a function of
incident neutrino energy for intrinsic charm only {($\langle k_T^2\rangle = 1.4$ GeV$^2$)}, for an 400 GeV proton beam with $2\times 10^{20}$ protons incident on a molybdenum target,
assuming a detector with a column depth of $L=66.7$ g/cm$^2$ of lead. }
\label{fig:events-ic}
\end{center}
\end{figure}

\begin{table}
\begin{centering}
\begin{tabular}{|c|c|c|c|}
\hline 
Production & $\nu_{\tau}$ & $\bar{\nu}_{\tau}$ & $\bar{\nu}_{\tau}+\nu_{\tau}$ \tabularnewline
\hline 
\hline 
pQCD & 188 (93, 262) & 84 (41, 116) & 272  (134, 378) \tabularnewline
\hline 
IC25 & 1379 & 606 & 1985 \tabularnewline
\hline 
IC7 & 386 & 170 & 556\tabularnewline
\hline 
\end{tabular}
\par\end{centering}
\caption{{The number of tau neutrino and antineutrino events at SHiP with $\eta_{\nu}>5.3$.
Evaluations are made from charm production and decay from a 400 GeV proton beam with $2\times10^{20}$
protons incident on a molybdenum target, with the neutrinos interacting 
in a lead detector with column depth $L=66.7\,\,\textrm{g/}\textrm{cm}^{2}$.  For the perturbative QCD evaluation (pQCD), the number of events
in parentheses are the numbers of events for the lower and upper limits of the shaded bands in fig. \ref{fig:events-scale}. The numbers for IC25
and IC7 are for intrinsic charm alone.}}
\label{tab:totalevents}
\end{table}

Intrinsic charm, at the level with normalization IC25 in eq. (\ref{eq:icnorm}), makes a large contribution to the overall number of events per GeV
of tau neutrinos (middle, orange histogram) and tau antineutrinos (lower, green histogram) shown in fig. \ref{fig:events-ic}.
The upper (blue) histogram is the sum of tau neutrino plus antineutrino events. The total number is {1985} events with these
parameter choices. If the normalization of the intrinsic charm contribution in $pN$ collisions 
is lowered such that $d\sigma/dx_F=7\ \mu$b {at $x_F=0.32$ (IC7)}, the number of events from
intrinsic charm reduces by a factor of 0.28. The double peak structure comes from the separate direct tau neutrinos
from the $D_s$ decay (lower energy peak) and the chain decay from $D_s\to\tau\to \nu_\tau$ with the higher energy peak.
Including smaller $\eta_\nu$ values (including larger angles around the beam direction) merges the two peaks. 
{A similar double peak structure is seen in the NLO perturbative QCD evaluation for $\eta_\nu>6.0$, but the peaks are overlapping 
for $\eta_\nu>5.3$, shown in fig. \ref{fig:events-scale}.}
{Again,} the
asymmetry between the numbers of tau neutrino and antineutrino events comes from the differences in the cross sections. 
The tau neutrino and antineutrino
fluxes are nearly identical for intrinsic charm, although this is not the case for muon neutrinos and antineutrinos.
{The number of events from tau neutrinos and antineutrinos with $\eta_\nu>5.3$ and interactions in the SHiP detector, modeled as a lead detector with column depth $L=66.7$ g/cm$^2$,  from a beam of $2\times 10^{20}$
protons with $E_p=400$ GeV incident on molybdenum are summarized in table \ref{tab:totalevents}.}
{We do not evaluate a detailed uncertainty for intrinsic charm here because the intrinsic charm contribution to forward charm production is only mildly
constrained. As remarked in sec. 2, the $A$ dependence uncertainty can be absorbed into the overall normalization. More detailed
analyses with fragmentation functions, binding energy and mass effects would be merited in the event that an intrinsic charm effect is measured.}

\begin{figure}
\begin{center}
\hspace{-0.85cm}
\includegraphics[width=0.75\textwidth]{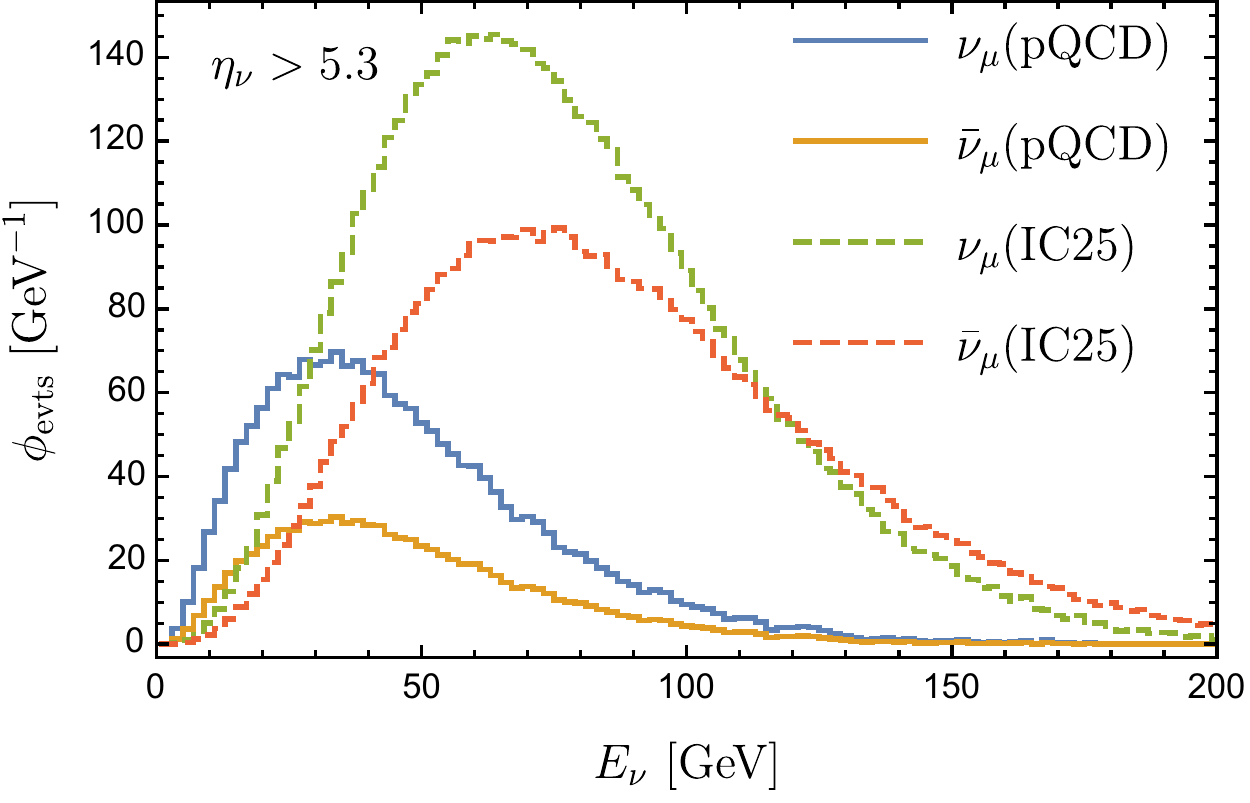}
\caption{The NLO perturbative QCD charm pair evaluation of the muon neutrino (solid blue) and muon antineutrino (solid orange) events per GeV for  $\eta_\nu>5.3$ as a function of
incident neutrino energy, for an 400 GeV proton beam with $2\times 10^{20}$ protons on target incident on a molybdenum target,
assuming a detector with a column depth of $L=66.7$ g/cm$^2$ of lead. Also shown are the intrinsic charm events per GeV where the intrinsic
charm cross section for $p$Mo is normalized with eq. (\ref{eq:icnorm}). }
\label{fig:events-muc}
\end{center}
\end{figure}

\begin{figure}
\begin{center}
\hspace{-0.85cm}
\includegraphics[width=0.7\textwidth]{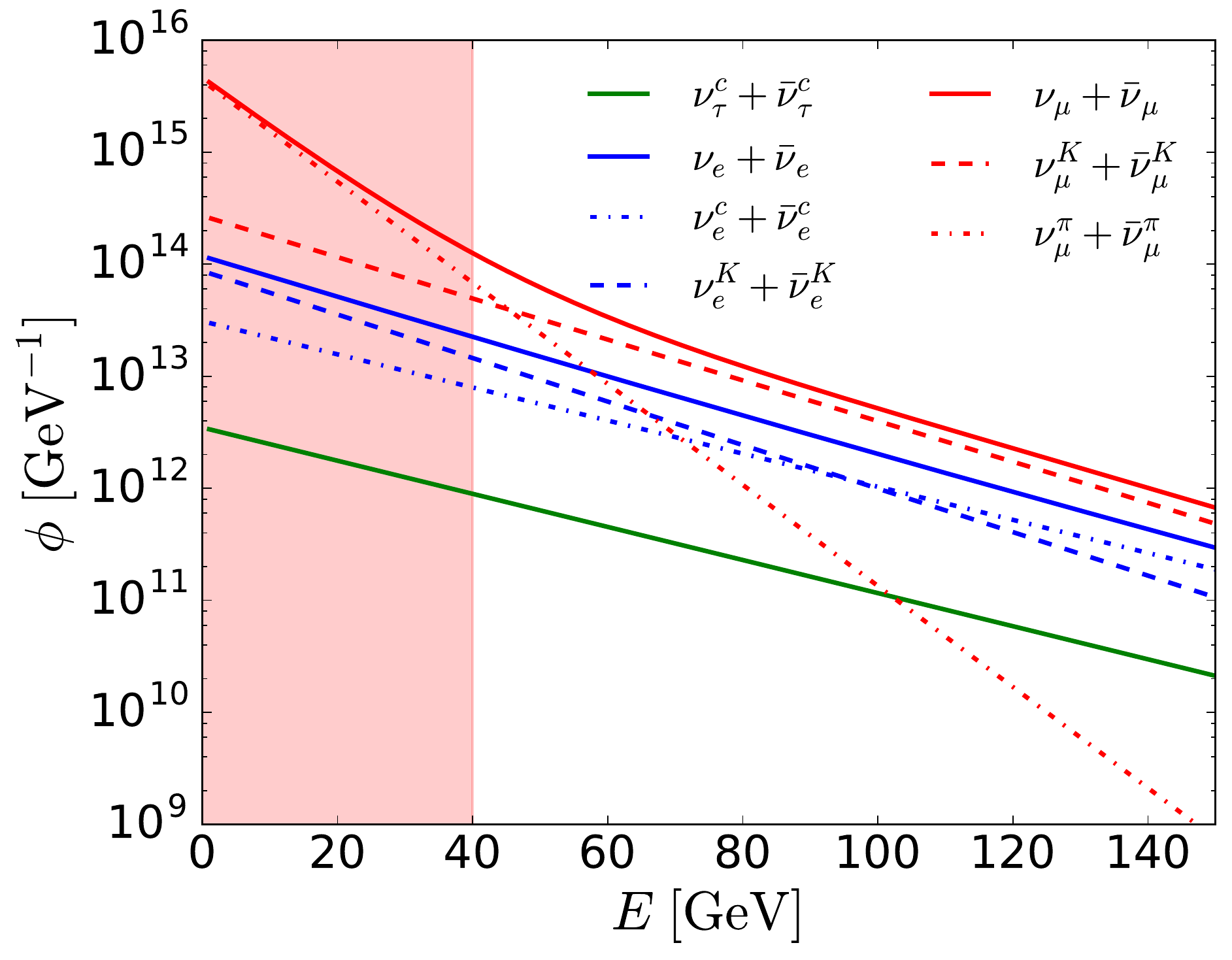}
\caption{Neutrino plus antineutrino fluxes (number per GeV) from   eq. (\ref{eq:nunubar}) to approximately describe the
SHiP fluxes evaluated by Monte Carlo in ref. \cite{Anelli:2015pba}, normalized by the $\nu_\tau +\bar{\nu}_\tau$ 
flux from perturbative charm production and decay (lower, green line).  Here, $\nu_e^c+\bar{\nu}_e^c=
\nu_\mu^c+\bar{\nu}_\mu^c$. The red shaded region indicates that the power law flux
is not reliable.}
\label{fig:conv}
\end{center}
\end{figure}

\begin{figure}
\begin{center}
\hspace{-0.85cm}
\includegraphics[width=0.7\textwidth]{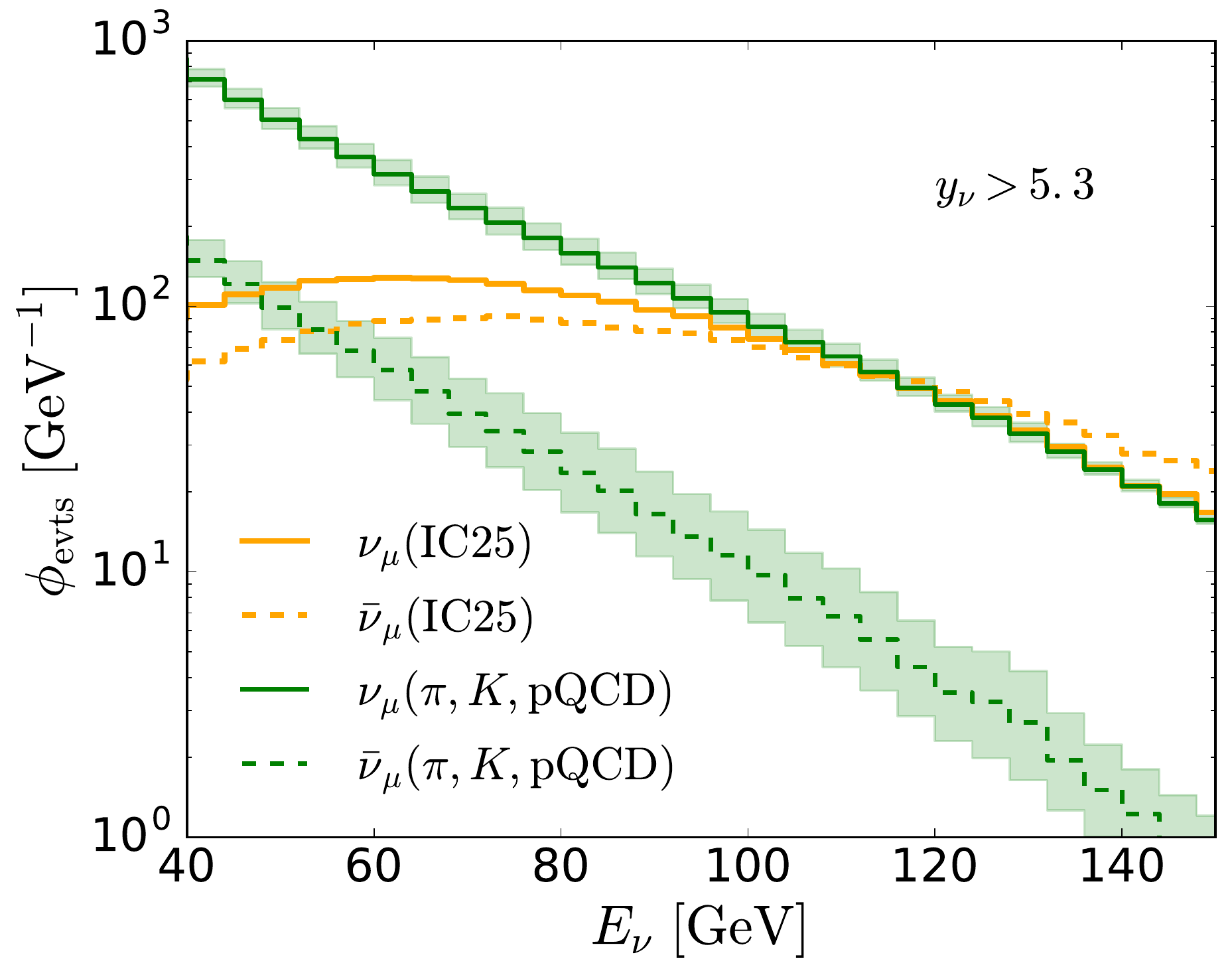}
\caption{The NLO perturbative QCD charm (pQCD) contribution of the muon neutrino (solid green) and muon antineutrino (dashed green) events for  $\eta_\nu>5.3$ added to the estimated pion and kaon contributions to muon neutrino and antineutrino events, as a function of
incident neutrino energy, for an 400 GeV proton beam with $2\times 10^{20}$ protons on target incident on a molybdenum target,
assuming a detector with $L=66.7$ g/cm$^2$ of lead. The shaded bands reflect an estimate of the range of uncertainty of a factor of 0.3-2 in the perturbative charm contributions. Also shown are the intrinsic charm events where the intrinsic
charm cross section for $p$Mo collisions is normalized with eq. (\ref{eq:icnorm}).   }
\label{fig:events-mu}
\end{center}
\end{figure}

We now turn to the number of muon neutrino and antineutrino events. We begin with NLO perturbative QCD predictions, including intrinsic transverse momentum and three-dimensional kinematics,
for muon neutrinos and antineutrinos from charm production and decay. Fig. \ref{fig:events-muc} shows the number
of muon neutrino and {antineutrino} events {per GeV} from charm as a function of energy 
for the 400 GeV proton beam with the standard parameter choices. The two dashed histograms show the intrinsic
charm number of events {per GeV}. 
While for the perturbative evaluation, the difference in neutrino and antineutrino number of events {per GeV} is due to the different
neutrino and antineutrino cross sections, the number of events {per GeV} for intrinsic charm reflects enhanced forward production of 
hadrons with a charm quark or antiquark that combines with valence quarks. Charm leptonic decays go to neutrinos, and
anticharm decays to an antineutrino. The meson components dominate over $\Lambda_c^\pm$, so the flux of antineutrinos
is higher than neutrinos. 

{For the perturbative evaluation of charm, the number of events from muon neutrino and antineutrino interactions at SHiP with 
$\eta_\nu>5.3$ are a factor of 20 larger than the
number of events from tau neutrino plus antineutrino interactions, a total of 5738 events for the sum of solid histograms 
(labeled pQCD) in fig. \ref{fig:events-muc}. The intrinsic charm IC25 number of events
for muon neutrinos and antineutrinos, a total of 21055 for the sum of dashed histograms in fig. \ref{fig:events-muc}, is a factor of $\sim 10$ larger than the number of events for tau neutrinos and antineutrinos from the IC25 evaluation.
Where there is no difference in the tau neutrino and antineutrino fluxes, the fraction of $\bar{\nu}_\tau$ events of the total number of $\nu_\tau+\bar{\nu}_\tau$ events is about 30\%, as is the fraction of 
$\bar{\nu}_\mu$ events to the total number of $\nu_\mu+\bar{\nu}_\mu$ events evaluated from perturbative QCD. The fraction of $\bar{\nu}_\mu $ to the total of $\nu_\mu+\bar{\nu}_\mu$ from IC25 alone is 44\%. }

\subsection{Muon charge asymmetry at SHiP}

The asymmetry in particle and antiparticle production with intrinsic charm is an interesting feature to 
use as a diagnostic for intrinsic charm. {As noted above, the asymmetry in charm hadron particle-antiparticle production will translate to the  numbers of $\nu_\mu$ and $\bar{\nu}_\mu$
per GeV, but not for $\nu_\tau$ and $\bar{\nu}_\tau$.}
While muon neutrinos and antineutrinos come from charm decays, they also come from pion and kaon decays, even in the beam dump environment. A full simulation of the muon neutrino plus antineutrino flux at SHiP is beyond the scope of this work, but we can
use the approximate flux evaluated in ref. \cite{Anelli:2015pba} for SHiP. The details of our normalization of their results (presented in ref. \cite{Anelli:2015pba} with arbitrary normalization) for the muon, electron and tau neutrino plus antineutrino fluxes
are discussed in appendix C. We review them briefly here.  Our strategy is to use the tau neutrino plus antineutrino flux to 
set the overall normalization. Assuming simple power laws and taking the electron neutrino plus antineutrino flux from charm, we can extract the contribution from kaons
to $\nu_e+\bar{\nu}_e$. 
The kaon contribution to $\nu_\mu+\bar{\nu}_\mu$ is approximated, and the pion contribution to the $\nu_\mu+\bar{\nu}_\mu$ 
flux is extracted. Charged particle production ratios \cite{Bonesini:2001iz} are used to approximate the separate neutrino and antineutrino fluxes. Fig. \ref{fig:conv} shows these fluxes with the separate pion, kaon and charm distributions. 
The flux of $\nu_\mu+\bar{\nu}_\mu$ from pions falls quickly with energy relative to those from kaons because of the 
relatively long pion lifetime. While the figure
shows an energy range to very small $E_\nu$, the simple power law certainly fails at low energy. Below 40 GeV neutrino energy, our approximate flux is less reliable {so we have shaded the region of $E_\nu<40$ GeV in fig. \ref{fig:conv}.}

\begin{figure}
\begin{center}
\hspace{-0.85cm}
\includegraphics[width=0.7\textwidth]{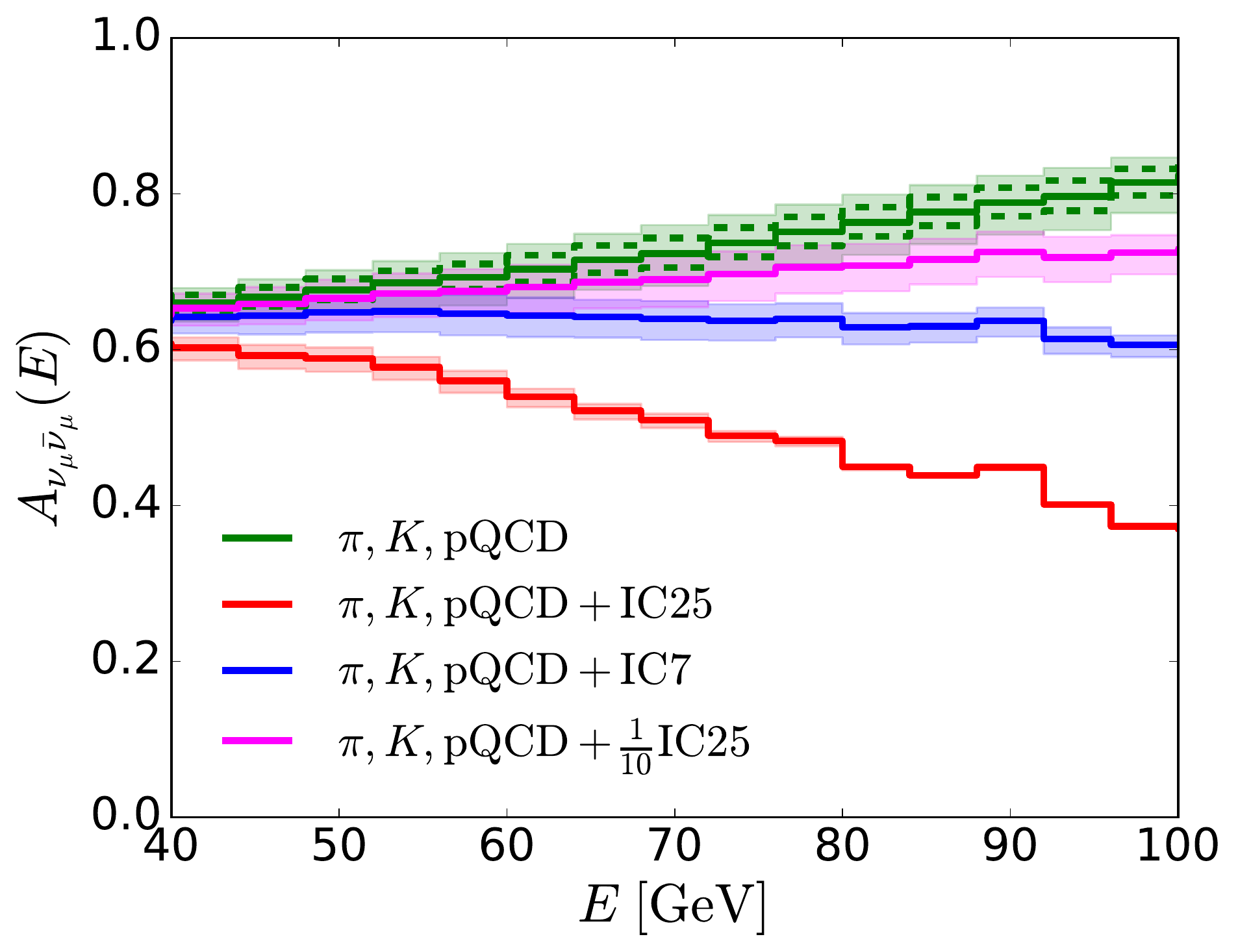}
\caption{Neutrino-antineutrino asymmetries in the number of events per GeV, as a function of energy, for muon neutrinos and antineutrinos,
as defined in eq. (\ref{eq:asym}). The dashed histograms reflect a $\pm 40\%$ factor multiplying the perturbative charm contribution to the asymmetry
without any intrinsic charm contribution. The shaded histogram bands show an estimate of the uncertainty in the perturbative charm production by
using factors of 0.30-2.0 of the central perturbative result for both the muon and antineutrino production.}
\label{fig:asymmetry}
\end{center}
\end{figure}

The number of muon neutrino and antineutrino events {per GeV} from pQCD production of charm combined with our modeled pion and kaon
neutrinos is shown in fig. \ref{fig:events-mu} with the green {histograms}: the upper solid {histogram} for $\nu_\mu$ charged current events per GeV and lower
dashed  green {histogram} for $\bar{\nu}_\mu$ events {per GeV} (no intrinsic charm). Low energy contributions come from pions,
however, for $\nu_\mu$, the kaon contribution dominates for $E$ larger than $\sim 40$ GeV. For antineutrinos, the crossover between
pion and kaon contributions is also at $\sim 40$ GeV, at which point the prompt contribution from charm is nearly equal to the kaon
contribution to the $\bar{\nu}_\mu$ {number of events per GeV}. 
{Since we have not done a full error analysis of the $D\to \nu_\mu$ modeling, we have expanded the error band for the perturbative
contributions to $\nu_\mu+\bar{\nu}_\mu$. The uncertainty in the total number of events for the perturbative QCD evaluation of the $\nu_\tau+\bar{\nu}_\tau$ events in 
table \ref{tab:totalevents}
is 0.5-1.4 times the central value. } The shaded bands reflect a {more} conservative error estimate for the perturbative 
muon neutrino and antineutrino fluxes, taking a factor of 0.3-2 times the central perturbative result. The green error band reflecting  charm
production of $\nu_\mu$
decreases with energy because at higher energies, the kaon contribution dominates. The green error band reflecting charm production of
$\bar{\nu}_\mu$ increases with energy as the charm contribution increases in importance with increasing energy.

The nearly flat orange {histograms} in fig. \ref{fig:events-mu} show the $\nu_\mu$ (solid) and $\bar{\nu}_\mu$ (dashed) number of events per GeV where they come from only intrinsic charm production. The intrinsic charm  events per GeV for $\nu_\mu$'s is approximately equal to that of muon neutrinos from the combined contributions from pions, kaons and perturbative charm, and for $\bar{\nu}_\mu$ from intrinsic charm, {all much larger than the $\pi$, $K$ and pQCD contributions to
the number of $\bar{\nu}_\mu$ events per GeV.}
One can define the charge asymmetry as
\begin{equation}
A_{\nu_\mu\bar{\nu}_\mu} = \frac{dN(\nu_\mu\to \mu^-)/dE-dN(\bar{\nu}_\mu\to\mu^+)/dE}{dN(\nu_\mu\to\mu^-)/dE + dN(\bar{\nu}_\mu\to\mu^+)/dE}\ ,
\label{eq:asym}
\end{equation}
{where $dN/dE = \phi_{\rm evts}(E)$ for the respective interactions.} Fig. \ref{fig:asymmetry} shows the charge asymmetry defined in eq. (\ref{eq:asym}).
The highest histogram shows the asymmetry evaluated with only pion, kaon and perturbative charm contributions to the $\nu_\mu+\bar{\nu}_\mu$
flux. The lowest curve shows the case where the intrinsic charm is normalized according to eq. (\ref{eq:icnorm}) for protons incident on molybdenum
(IC25) added to the perturbative charm, pion and kaon contributions. Intermediate curves show the results {for IC7 (eq. (\ref{eq:ic7}))
and when the IC25 normalization is decreased}
by a factor of 1/10. The error band shown with the dashed histogram for $\pi,K$ and perturbative charm production of muon neutrinos and
antineutrinos shows an approximate {energy independent} $\pm 40\%$ error on the perturbative $\nu_\mu +\bar{\nu}_\mu$ flux that comes from scale dependence.
The shaded bands show 
{a conservative} error estimate of the perturbative error with a range of 0.3-2.0 times the central perturbative $\nu_\mu +\bar{\nu}_\mu$ result.
{Even with the larger error band on the perturbative contributions,  the change in the muon charge asymmetry when intrinsic charm
contributes at the level of the IC25 or IC7 normalization is evident.}
A full study of perturbative and nonperturbative contributions to the $\nu_\mu$ and $\bar{\nu}_\mu$ flux is beyond the scope of this work.
However, with our approximations for the pion and kaon contributions to the number of $\nu_\mu+\bar{\nu}_\mu$, it is our estimate that the number of intrinsic charm events normalized according to IC7 or larger will be reflected in a muon charge asymmetry different from the results from 
pion, kaon and perturbative charm contributions to that asymmetry. 

\section{Discussion}

Tau neutrino and antineutrino detection at SHiP would show not just evidence of tau neutrino charged current production of taus, but also allow for
the study of tau neutrino interactions and the details of tau neutrino production. 
By approximating the detector acceptance for the SHiP detector a distance of 51.5m downstream from the front of the target with an angular cut
equivalent to $\eta_\nu>5.3$, we can reproduce our earlier results in ref. \cite{Alekhin:2015byh} to within 10\%.  Our  new
NLO perturbative QCD evaluation of
charm mesons, in particular for $D_s$ that includes intrinsic transverse momentum effects in the hard scattering and the angular effects in the decays of the $D_s\to \tau\nu_\tau$ and $\tau \to \nu_\tau X$, reduces the number of tau
events by a factor of $\sim 3$ compared to the number of events in ref. \cite{Alekhin:2015byh}, predicting {272} events with our central 
renormalization and factorization scale choices {and $\langle k_T^2\rangle = 1.4$ GeV$^2$. Scale and intrinsic transverse momentum
uncertainties introduce
a $\sim \pm 30-40\%$ } uncertainty in the number of events. {A larger charm quark mass or scale choice that depends on $m_T$ rather than $m_c$ results
in a smaller number of events. Combining mass, scale factor and other uncertainties leads to a lower value of {134 events and an upper value of
378 events, as summarized in table 3. The wide range of perturbative QCD predictions for the number of tau neutrino events for $\eta_\nu>5.3$,
here estimated to be a factor of 2.8,
emphasizes the importance of additional experimental measurements of charm productions, for example,  with the proposed DsTau experiment \cite{Aoki:2017spj}.}
}

Forward production of charm that is not perturbative can enhance neutrino production in beam dump experiments. We have used the BHPS
model for intrinsic charm \cite{Brodsky:1980pb,Brodsky:1981se,Vogt:1994zf,Gutierrez:1998bc} as a representative model to show the impact that
forward charm production can have on the number of tau neutrino and antineutrino events. With an intrinsic charm normalization suggested in the 
context of the prompt atmospheric neutrino flux, which also emphasizes forward production, we find that intrinsic charm production and decay to 
tau neutrinos and antineutrinos would significantly enhance the number of events. Indeed, intrinsic charm is relatively more important at low energies, e.g.,
at SHiP,  than for high energy {cosmic ray} interactions {in the atmosphere}
because the cross section is assumed to scale like the total hadronic cross section, so {it is} weakly dependent on energy, unlike the perturbative charm cross section. With the higher intrinsic charm normalization suggested in ref. \cite{Laha:2016dri} {(IC25)},
the number of events for tau production from intrinsic charm alone is almost {2,000} tau neutrino plus antineutrino charged current events.
{The energy distribution of the number of tau neutrino and antineutrino events, 
different for perturbative production of charm and for intrinsic charm, will be essential to untangle the
their relative contributions given the uncertainties in the total number of events.}

{Beyond the enhancement in the number of events,} one of the important features of intrinsic charm models is the asymmetry in particle production because of the valence momenta in the proton beam. The valence quarks are not relevant to $D_s^\pm$ production, so particle-antiparticle asymmetries {different from the NLO perturbative QCD predictions} won't appear in the tau neutrino and tau antineutrino events, but they will appear in muon neutrino and antineutrino charged current events. To estimate the $\mu^- - \mu^+$ asymmetry as a function of
energy, we used the Monte Carlo evaluation of the neutrino fluxes from ref. \cite{Anelli:2015pba}, then used particle production asymmetries 
\cite{Bonesini:2001iz} to estimate the separate pion and kaon contributions to the muon neutrino and antineutrino fluxes. Signals of intrinsic charm would be offsets in the particle asymmetry predicted from just pion and kaon contributions to the muon neutrino and antineutrino fluxes.

We have used the BHPS model for intrinsic charm to compare with what has been used to evaluate prompt atmospheric neutrino fluxes in ref. \cite{Laha:2016dri}, however, this is not the only model with forward charm production or for intrinsic charm. Within the BHPS model, {we have chosen simplified fragmentation in the uncorrelated fragmentation and coalescence contributions  and equal weights for the two contributions.}  These approximations can be changed {and refinements to the treatment of fragmentation can be made}. Their impacts on predictions can be assessed should experimental data point to intrinsic charm.

Presented here are neutrino fluxes with a single proton interaction, responsible for the bulk of the high energy neutrinos. Secondary and tertiary interactions will produce more, lower energy, neutrinos to increase the number of events. 
A full Monte Carlo of pion, kaon and charm hadron production with multiple interactions within the beam dump target is beyond the scope of this work. Ideally, a full simulation would be performed and compared to direct measurements of charm hadron production. 

In lieu of the next-to-leading order QCD evaluation of charm pair production with additional intrinsic transverse momentum, the Pythia Monte Carlo can be used to model charm production. A disadvantage of Pythia is that it is tuned to central production rather than forward production. Comparisons of Pythia modeling of charm production with the
E791 results in
$\pi^- N$ collisions with a 500 GeV beam energy \cite{Aitala:1999yp} showed that the data do not match well with the default tune, as noted in ref. \cite{Dijkstra:2015note}.  Dijkstra and Ruf, in ref. \cite{Dijkstra:2015note} use a retuned Pythia and consider multiple interactions within the target to produce charm. They find that the total amount of charm produced is 2.3 times the amount produced in the primary $pN$ interaction, although, with a softer spectrum, so the number of events will not increase by as large a factor.

Neutrino events in a beam dump experiment like SHiP will provide important tests of lepton flavor universality in
tau neutrino and antineutrino interactions. 
Both tau and muon flavors of neutrino
and antineutrino number of events, as a function of neutrino energy, will enable a disentangling of perturbative and non-perturbative elements of charm production. The small angular region around the proton beam direction required for neutrinos to intercept the SHiP detector probes very forward
charm production, complementary to the higher energy measurements of charm in collider experiments. The high pseudorapidity and relatively
low beam energy of SHiP will permit stronger constraints
on models of intrinsic charm, with their normalizations only weakly dependent on energy, than at high energy colliders.
Better measurements of charm production, either directly or through their neutrino decays, will help narrow uncertainties for 
neutrino fluxes from  charm produced by cosmic ray interactions in the atmosphere, a background to the diffuse flux of astrophysical neutrinos.

\acknowledgments

This work is supported in part by a US Department of Energy grant DE-SC-0010113. MHR acknowledges discussions with F. Tramontano and 
collaboration with Y. S. Jeong for the development of the collinear approximated fluxes.

\appendix
\section{Intrinsic charm probability distributions}

For completeness, we include the differential distributions of the $x_F$ distributions for intrinsic charm production in the BHPS model described in detail by Gutierrez and Vogt in ref. \cite{Gutierrez:1998bc}.
The $x_F$ probability distributions depend on an uncorrelated fragmentation function (F) and coalescence distributions (C) that are specific for
each charm hadron produced. We consider incident protons, where the valence quarks together with $Q\bar{Q}=c\bar{c}$ fluctuations contribute
to the 5 quark configuration, and an additional $q\bar{q}=u\bar{u},\ d\bar{d},\ {s\bar{s}}$ pair contributes to 7 quark configurations. As outlined in the 
appendix of ref. \cite{Gutierrez:1998bc} for $D^\pm$, $D_s^\pm$ and $\Lambda_c^\pm$, and extended to $D^0$, $\bar{D}^0$, the probability distributions are
\begin{eqnarray}
\nonumber
\frac{dP_{D^-}}{dx_F} & = &
                        \frac{1}{2}\Biggl(\frac{1}{4}\frac{dP_{ic}^{5C}}{dx_F}
+ \frac{1}{10}\frac{dP_{ic}^{5F}}{dx_F}\Biggr) 
+    \frac{1}{2}\Biggl(\frac{1}{5}\frac{dP_{icu}^{7C}}{dx_F}
+ \frac{1}{10}\frac{dP_{icu}^{7F}}{dx_F}\Biggr)\\
& + &
 \frac{1}{2}\Biggl(\frac{2}{5}\frac{dP_{icd}^{7C}}{dx_F}
+ \frac{1}{10}\frac{dP_{icd}^{7F}}{dx_F}\Biggr)
+                        \frac{1}{2}\Biggl(\frac{1}{5}\frac{dP_{ics}^{7C}}{dx_F}
+ \frac{1}{10}\frac{dP_{ics}^{7F}}{dx_F}\Biggr)\\
\nonumber
\frac{dP_{D^+}}{dx_F} & = &
                       \frac{1}{10}\frac{dP_{ic}^{5F}}{dx_F}
+ \frac{1}{10}\frac{dP_{icu}^{7F}}{dx_F}
+                       \frac{1}{2}\Biggl(\frac{1}{8}\frac{dP_{icd}^{7C}}{dx_F}
+ \frac{1}{10}\frac{dP_{icd}^{7F}}{dx_F}\Biggr)
+ 
\frac{1}{10}\frac{dP_{ics}^{7F}}{dx_F}\\ 
\nonumber
\frac{dP_{\bar{D}^0}}{dx_F} & = &
                        \frac{1}{2}\Biggl(\frac{1}{2}\frac{dP_{ic}^{5C}}{dx_F}
+ \frac{1}{10}\frac{dP_{ic}^{5F}}{dx_F}\Biggr)
+                        \frac{1}{2}\Biggl(\frac{3}{5}\frac{dP_{icu}^{7C}}{dx_F}
+ \frac{1}{10}\frac{dP_{icu}^{7F}}{dx_F}\Biggr)\\
\nonumber
& + &
                        \frac{1}{2}\Biggl(\frac{2}{5}\frac{dP_{icd}^{7C}}{dx_F}
+ \frac{1}{10}\frac{dP_{icd}^{7F}}{dx_F}\Biggr) 
              +          \frac{1}{2}\Biggl(\frac{2}{5}\frac{dP_{ics}^{7C}}{dx_F}
+ \frac{1}{10}\frac{dP_{ics}^{7F}}{dx_F}\Biggr)\\
\nonumber
\frac{dP_{D^0}}{dx_F} & = &
                       \frac{1}{10}\frac{dP_{ic}^{5F}}{dx_F}
+ \frac{1}{10}\frac{dP_{icd}^{7F}}{dx_F}
          +             \frac{1}{2}\Biggl(\frac{1}{8}\frac{dP_{icu}^{7C}}{dx_F}
+ \frac{1}{10}\frac{dP_{icu}^{7F}}{dx_F}\Biggr)
+ 
\frac{1}{10}\frac{dP_{ics}^{7F}}{dx_F}
\end{eqnarray}
\begin{eqnarray}
\nonumber
\frac{dP_{D_s^-}}{dx_F} & = &
                        \frac{1}{10}\frac{dP_{ic}^{5F}}{dx_F}
+ \frac{1}{10}\frac{dP_{icu}^{7F}}{dx_F} + \frac{1}{10}\frac{dP_{icd}^{7F}}{dx_F}  
 + 
 \frac{1}{2}\Biggl(\frac{1}{5}\frac{dP_{ics}^{7C}}{dx_F}
+ \frac{1}{10}\frac{dP_{ics}^{7F}}{dx_F}\Biggr)\\
\nonumber
\frac{dP_{D_s^+}}{dx_F} & = &
                       \frac{1}{10}\frac{dP_{ic}^{5F}}{dx_F}
+ \frac{1}{10}\frac{dP_{icu}^{7F}}{dx_F}
+                       \frac{1}{10}\frac{dP_{icd}^{7F}}{dx_F}
+ \frac{1}{2}\Biggl(
\frac{1}{8}\frac{dP_{ics}^{7C}}{dx_F}+ \frac{1}{10}\frac{dP_{ics}^{7F}}{dx_F}\Biggr) \\
\nonumber
\frac{dP_{\Lambda_c^-}}{dx_F} & = &
                        \frac{1}{10}\frac{dP_{ic}^{5F}}{dx_F}
+ \frac{1}{10}\frac{dP_{icu}^{7F}}{dx_F} + \frac{1}{10}\frac{dP_{icd}^{7F}}{dx_F}  
+ \frac{1}{10}\frac{dP_{ics}^{7F}}{dx_F}\\
\nonumber
\frac{dP_{\Lambda_c^+}}{dx_F} & = & \frac{1}{2}\Biggl(
                      \frac{1}{2} \frac{dP_{ic}^{5C}}{dx_F} + \frac{1}{10}\frac{dP_{ic}^{5F}}{dx_F} \Biggr)
+ \frac{1}{2}\Biggl( \frac{3}{8}\frac{dP_{icu}^{7C}}{dx_F}+\frac{1}{10}\frac{dP_{icu}^{7F}}{dx_F} \Biggr)\\ 
\nonumber
&+&                      \frac{1}{2}\Biggl( \frac{1}{2}\frac{dP_{icd}^{7C}}{dx_F}  +\frac{1}{10}\frac{dP_{icd}^{7F}}{dx_F} \Biggr)
+ \frac{1}{2}\Biggl(
\frac{1}{4}\frac{dP_{ics}^{7C}}{dx_F}+ \frac{1}{10}\frac{dP_{ics}^{7F}}{dx_F}\Biggr) 
\end{eqnarray}
If the sum of charged and neutral $D$'s (but not $D_s^\pm$ and $\Lambda_c^\pm$) for $pN$ interactions gives $d\sigma/dx_F(x_F=0.32)=25\ \mu$b, the
cross sections are those shown in table \ref{table:icsigma}. 

\begin{table}[tb]
\begin{center}
\begin{tabular}{|l|c|c|c|}
\hline
Particle H & $\sigma_{ic}^{\rm H}\ [\mu$b]  & Particle H & $\sigma_{ic}^{\rm H}\ [\mu$b] \\
\hline
$D^0$ & 2.14  & $\overline{D}^0$ & 4.71\\
\hline
$D^+$ & 2.16 & $D^-$ & 3.09   \\
\hline
$D_s^+$ & 2.19  &$D_s^-$ & 2.28 \\
\hline
$\Lambda_c^+ $ & 4.37 & $\Lambda_c^-$ & 2.27\\
\hline
\end{tabular}
\caption{\label{table:icsigma} 
Cross section for $pN$ interactions, where the normalization is $d\sigma/dx_F=25\ \mu$b at $x_F=0.32$ for the sum of $D^0$, $\overline{D}^0$,
$D^+$ and $D^-$ distributions. For $pA$ interactions, the cross section is multiplied by $A^{0.71}$.}
\end{center}
\vspace{-0.6cm}
\end{table}

\section{Charm hadron decays to neutrinos}
\label{sec:ctonu}

Tau neutrinos and antineutrinos come from $D_s$ production and their decay
$D_s^+\to \tau^+ \nu_\tau$ (direct neutrino) and tau decay $\tau^+\to\bar{\nu}_\tau X$
(chain decay neutrino), and charge conjugate processes. The branching fraction is
$B(D_s\to \tau \nu_\tau)= (5.55\pm 0.23)\%$ \cite{Patrignani:2016xqp}. We have implemented the $D_s^\pm$ and $\tau^\pm$ decay
distributions {into the HVQ} computer code. We sketch some of the inputs here.
The direct and chain decay neutrino energy distributions from $\pi^\pm$ and $\mu^\pm$ decays are discussed in
detail in refs. \cite{Barr:1988rb,Barr:1989ru,gaisser1990cosmic}. The evaluation is directly transferable
to $D_s\to \tau\nu_\tau$. 

In the $D_{s}$ rest frame,
$\nu_\tau$'s are produced isotropically since the $D_s$ has zero spin. For this two-body
decay, in the rest frame of the $D_s$, the tau and neutrino energies are
\begin{eqnarray}
E_{\tau}^{'}&=&\frac{m_{D_{s}}^{2}+m_{\tau}^{2}}{2m_{D_{s}}}=\frac{1}{2}m_{D_{s}}(1+r)=1.786\ {\rm GeV}\nonumber \\
 E_{\nu_{\tau}}^{'}&=&\frac{m_{D_{s}}^{2}-m_{\tau}^{2}}{2m_{D_{s}}}=\frac{1}{2}m_{D_{s}}(1-r)=0.182\ {\rm GeV}\ ,
\end{eqnarray}
where $r=(m_{\tau}/m_{D_s})^{2}=0.815$, $m_{D_{s}}=1.97$ GeV and
$m_{\tau}=1.78$ GeV. We use primed variables to denote quantities
in the $D_{s}$ rest frame. One can find the corresponding distributions in the lab frame through boosts 
which we evaluate numerically. Ref. \cite{McDonald:2001mc} has a discussion of the details of the boosts, for example.
The $\nu_\tau$'s in the two-body $D_s$ decays are the direct neutrinos.

To get the energy and momentum of the chain decay neutrinos, we begin with the tau in its rest frame.
In the tau rest frame, the distribution of the massless (or nearly massless)
leptons from the three-body decay of the tau is given by \cite{Barr:1988rb,Barr:1989ru,gaisser1990cosmic}
\begin{equation}
\frac{\mathrm{d}n^{*}}{\mathrm{d}x\mathrm{^{*}d}\Omega^{*}}=\frac{1}{4\pi}\left[f_{0}(x^{*})\mp f_{1}(x^{*})\cos\theta^{*}\right]
\label{eq:cth}
\end{equation}
where $x^{*}\equiv{2E_{l}^{*}}/{m_{\tau}}\in[0,1]$,  and $\theta^{*}$
is the angle between the lepton and the spin of the tau. For $\nu_{\tau}$
or $\overline{\nu}_{\tau}$, $f_{0}(x^{*})=2x^{*2}(3-2x^{*})$ and
$f_{1}(x^{*})=2x^{*2}(1-2x^{*})$. The starred variables denote the tau rest frame. Similar results apply to other tau decay modes.

The tau decay distribution in its rest frame, for angles of lepton $l=\nu_\tau (\bar{\nu}_\tau)$ or $D_s$ momenta relative to an axis in the lab direction of the tau momentum, can be rewritten as
\begin{eqnarray}
\frac{\mathrm{d}n^{*}}{\mathrm{d}x^{*}\mathrm{d}\Omega^{*}} & =&\frac{1}{4\pi}\left[f_{0}(x^{*})\mp f_{1}(x^{*})\cos\theta^{*}\right]\nonumber \\
 & =&\frac{1}{4\pi}\left[f_{0}(x^{*})-f_{1}(x^{*})\left(\cos\theta_{D_{s}}^{*}\cos\theta_{l}^{*}+\sin\theta_{D_{s}}^{*}\sin\theta_{l}^{*}\cos\phi_{l}^{*}\right)\right]\ ,
 \label{eq:neutrinoDisMuonRest}
\end{eqnarray}
where the opposite helicities of $\nu_\tau$ and $\bar{\nu}_\tau$ conspire to yield a single equation for both neutrino and antineutrino distributions, with
\cite{Barr:1988rb,Barr:1989ru}
\begin{equation}
\cos\theta_{D_{s}}^{*}=\frac{1}{\beta_{\tau}}\left(\frac{2E_{D_{s}}r}{E_{\tau}\left(1-r\right)}-\frac{1+r}{1-r}\right)\equiv P_{\tau}.\label{eq:polarization-1}
\end{equation}
where $E_{D_{s}}$ and $E_{\tau}$ are the energies of $D_{s}$
and $\tau$ in the lab frame. 
The results shown below use eq. (\ref{eq:neutrinoDisMuonRest}) rotated and boosted to the lab frame for the tau
leptonic decay. The remainder of the tau decays are approximated by two-body decays, to $\pi \nu_\tau$,
$\rho \nu_\tau$ and $a_1\nu_\tau$. The decay distributions can be found in ref. \cite{Bhattacharya:2016jce}.

As noted above, particle-antiparticle asymmetries in charm hadron production will not be evident for $D_s^\pm$ but will be for other $D$ mesons and the 
$\Lambda_c$. To estimate signals in SHiP, we consider the decays $H\to \nu_\mu X$. 
For the muon neutrino and antineutrino distributions shown here, we have used an approximate energy distribution equal to the muon energy distribution. As implemented in the HVQ program, the parameterization of the muon distribution follows a naive spectator model which
depends on the mass ratio $r=m_K^2/m_H^2$, where $m_H$ is the charm hadron mass and $m_K$ is the effective mass of the hadronic final state in the decay.

\section{Neutrino flux at SHiP from pions and kaons}
\label{sec:pik}

\begin{figure}
\begin{center}
\hspace{-0.85cm}
\includegraphics[width=0.75\textwidth]{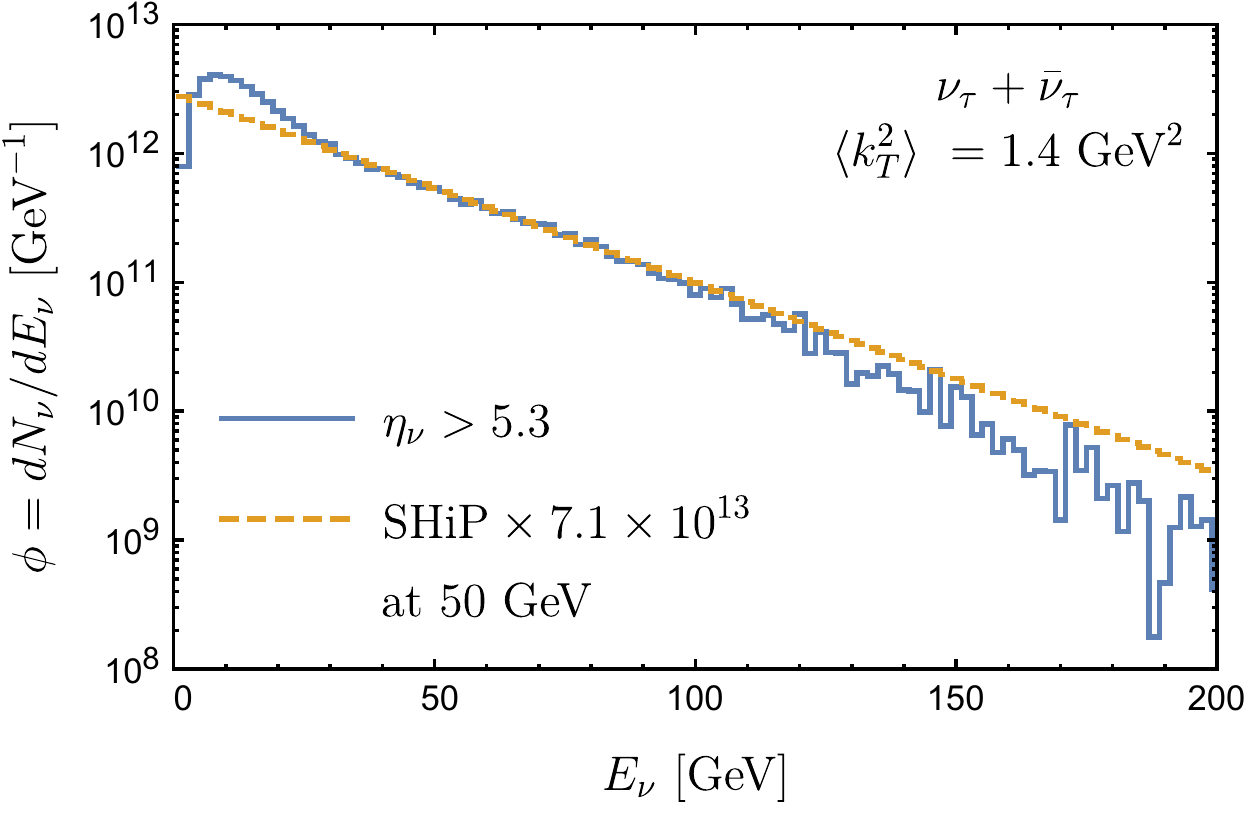}
\caption{Tau neutrino plus antineutrino flux evaluated using NLO QCD (histogram) and with an approximate curve described by eq. (\ref{eq:nunubar}) to approximately describe using a power law the tau neutrino plus antineutrino flux (arbitrary normalization shown in ref. \cite{Anelli:2015pba}).}
\label{fig:tau-c-norm}
\end{center}
\end{figure}

The muon neutrino-antineutrino asymmetry has contributions from pions and kaons that decay to neutrinos since $pA$ production ratios of positively charged mesons to negatively charged pions and kaons is not unity. A full evaluation of the neutrino
flux from pions and kaons is beyond the scope of this work, but we estimate their contributions to the electron and muon neutrino fluxes and muon asymmetry using the flux estimation of ref. \cite{Anelli:2015pba}.
In fig. 5.25 of ref. \cite{Anelli:2015pba}, the neutrino fluxes into a detector 56.5 m from the target, with an
area of 1.3 m$^2$, are shown from an evaluation that includes the full kinematics from $D_s$ decays and a GEANT4 evaluation of pion and kaon production by 400 GeV protons incident on a molybdenum target. Their figure is normalized to 100 neutrinos, and it is approximated by the solid curves shown in fig.
\ref{fig:conv}, except for the low energy fluxes. We use our evaluation of the high energy
flux of $\nu_\tau+\bar{\nu}_\tau$ to normalize the flux of neutrinos from pions and kaons in ref. \cite{Anelli:2015pba}, as described below. The tau neutrino flux is
approximately
\begin{equation}
\label{eq:nunubar}
\phi^c_{\nu_\tau+\bar{\nu}_\tau} = 4.0\times 10^{-2}\, N_\nu\, \exp\Bigl[ -E/29.4 \ {\rm GeV}\Bigr]\ .
\end{equation}
The tau neutrino plus antineutrino flux determines the overall normalization constant to be {approximately}
$N_\nu=7.1\times 10^{13}$ neutrinos/GeV, by matching our perturbative $\phi_{\nu_\tau+\bar{\nu}_\tau}$ with eq. 
(\ref{eq:nunubar}) for  energies between $E\simeq 50-100$ GeV and $\eta_\nu>5.3$. The comparison of our flux evaluation of tau neutrinos plus antineutrinos from perturbative charm production and the normalized curve from eq. (\ref{eq:nunubar}) is shown in fig. \ref{fig:tau-c-norm}.

Given the flux of $\nu_\tau+\bar{\nu}_\tau$, from charm, we can subtract the charm contributions from the total $\nu_e+\bar{\nu}_e$ and
$\nu_\mu+\bar{\nu}_\mu$ fluxes. We take the ratio of the electron neutrino flux at $E_\nu=200$ GeV to the tau neutrino flux at the same energy equal to a
rescaling constant to relate the charm contributions to the fluxes $\phi^c_{\nu_e+\bar{\nu}_e}=
\phi^c_{\nu_\mu+\bar{\nu}_\mu}$ to $\phi_{\nu_\tau+\bar{\nu}_\tau}$, since the tau neutrino flux only comes from charm.  The high energy contributions of kaons (and pions to $\nu_\mu+\bar{\nu}_\mu$) is negligible due to meson
reinteractions. In
fig. \ref{fig:conv}, extrapolated to 200 GeV, the ratio is $N_c=8.9$. The flux $\phi^c_{\nu_e+\bar{\nu}_e}=
\phi^c_{\nu_\mu+\bar{\nu}_\mu}$ is shown with the (blue) dot-dashed line.

The difference between the total neutrino fluxes and the charm contributions gives the contributions from pions and kaons. Since pions don't contribution to $\nu_e+\bar{\nu}_e$, the difference in curves is due only to kaons. After the charm subtraction, then, 
\begin{eqnarray} 
\phi^K_{\nu_e+\bar{\nu}_e}&=&\phi_{\nu_e+\bar{\nu}_e}-
N_c\phi_{\nu_\tau+\bar{\nu}_\tau}\\
\phi^K_{\nu_\mu+\bar{\nu}_\mu}+\phi^\pi_{\nu_\mu+\bar{\nu}_\mu}&=&\phi_{\nu_\mu+\bar{\nu}_\mu}-
N_c\phi_{\nu_\tau+\bar{\nu}_\tau}\ .
\end{eqnarray}
The kaon contribution to the electron neutrino flux is shown with the dashed blue curve in the figure.

To determine the kaon contribution to the muon neutrino flux, the ratio of charged to neutral kaons is needed.
Kaon production is taken in proportions  
$N(K_S)=N(K_L)=(N(K^+)+3 N(K^-))/4$ according to an approximation of valence
$u_v=2d_v$ and sea $u_s=\bar{u}
=d_s=\bar{d}=s=\bar{s}$  in a simple isospin symmetric model of production \cite{Bonesini:2001iz}.
The ratios of positive to negative particle production, in terms of the ratio $x_R= E^*/E^*_{max}$, the
ratio of the energy of the kaon (and for reference, the pion) in the center of momentum frame to the maximum energy,
approximately equal to the ratio of the energies in the lab frame, are \cite{Bonesini:2001iz,Ochs:1976jh}
\begin{eqnarray}
\label{eq:ratpi}
r_\pi (x_R)&=& 1.05\cdot (1+x_R)^{2.65}\\
\label{eq:ratk}
r_K(x_R) &=& 1.15\cdot(1-x_R)^{-3.17}\ .
\end{eqnarray}
The charge ratio for pions saturates around $\sim 5$, while the charge ratio for kaons increases with kaon energy
\cite{Ochs:1976jh}, which translates to more neutrinos than antineutrinos from pions and kaons.
The charge ratios and branching fractions to $\nu_e$ and $\nu_\mu$ allow for an approximate energy
dependent scaling of the kaon contribution to the electron neutrino flux already determined, 
to the kaon contribution to the 
muon neutrino flux. The kaon contribution to the muon neutrino flux is shown with the upper
(red) dashed line in fig. \ref{fig:conv}. For $x_R$ in eq. (\ref{eq:ratk}), we have made the approximation that $E_\nu\sim (1/3-1/2)E_K$ to find the energy dependent ratio between electron and muon neutrino fluxes for the dominant decays: $K^0_{e3}$, $K^0_{\mu 3}$, $K^+_{e 3}$, $K^+_{\mu 3}$, and $K^+\to
\mu^+\nu_\mu$. 
The muon neutrino plus antineutrino flux from pion production, with its subsequent decay to $\pi^+\to \mu^+\nu_\mu$, with a positive to negative charge ratio from eq. (\ref{eq:ratpi}), is shown in fig. \ref{fig:conv} with the steeply falling (red)
dot-dashed line.

The curves shown in fig. \ref{fig:conv}, with the overall  normalization, $N_\nu=7.1\times 10^{13}$ neutrinos/GeV, can
be parameterized by (see also eq. (\ref{eq:nunubar})):
\begin{eqnarray}
\phi^c_{\nu_\mu+\bar{\nu}_\mu} &=& \phi^c_{\nu_e+\bar{\nu}_e} = 8.9\, \phi_{\nu_\tau+\bar{\nu}_\tau} \\
\phi^K_{\nu_e+\bar{\nu}_e} &=&  N_\nu\, \exp\Bigl[ -E/22.4 \ {\rm GeV}\Bigr]\\
\label{eq:mumubark}
\phi^K_{\nu_\mu+\bar{\nu}_\mu} &=& N^K(E_\nu) \times \phi^K_{\nu_e+\bar{\nu}_e}\\
\label{eq:mumubarpi}
\phi^\pi_{\nu_\mu+\bar{\nu}_\mu} &=& 50.0\, N_\nu\, \exp\Bigl[ -E/9.63\ {\rm GeV}\Bigr]\ .
\end{eqnarray}
The relative normalizations between the kaon contributions to the muon neutrino and electron neutrino fluxes depend
on $r_K$ via
\begin{eqnarray}
\label{eq:knorm}
N^K(E_\nu) &=& \frac{ 0.27 N_{K_L} +0.64\bigl(N_{K^+}(2 x_\nu)+N_{K^-}(2 x_\nu)\bigr)}{0.41 N_{K_L} + 0.051 \bigl(N_{K^+}(3x_\nu)
+ N_{K^-}(3x_\nu)\bigr)}\\
N_{K^+}(x_R) &=& r_K(x_R) \, N_{K^-}\\
N_{K^-}&=& \frac{4}{3+r_K(x_R)}\, N_{K_L}
\end{eqnarray}
where $x_\nu= E_\nu/E_p$. For kaons, we take the energy fraction carried by the muon neutrino to be on average a factor of 1/2 of the kaon energy, while for the electron neutrino, we use a factor of 1/3, but the results are not very sensitive to this choice. The numerical factors in eq. (\ref{eq:knorm}) come from the dominant branching fractions to muons and electrons in kaon decays.
To determine the asymmetry in muon neutrino and muon antineutrino events, we use the overall fluxes from pions and kaons in eqs. (\ref{eq:mumubark}-\ref{eq:mumubarpi}) and the charge ratios
in eqs. (\ref{eq:ratpi}-\ref{eq:ratk}).

\providecommand{\href}[2]{#2}\begingroup\raggedright\endgroup

\end{document}